\documentclass[universe,article,accept,pdftex,moreauthors]{Definitions/mdpi} 
\usepackage{graphicx}
\usepackage{multirow}
\usepackage{amsmath,amssymb,amsfonts}
\usepackage{amsthm}
\usepackage{mathrsfs}
\usepackage[title]{appendix}
\usepackage{xcolor}
\usepackage{textcomp}
\usepackage{manyfoot}
\usepackage{booktabs}
\usepackage{algorithm}
\usepackage{algorithmicx}
\usepackage{algpseudocode}
\usepackage{listings}
\usepackage{aas_macros}
\usepackage{caption}
\usepackage{subcaption}

\newcommand{\be}{\begin{equation}}
\newcommand{\ee}{\end{equation}}

\newcommand{\unit}[1]{\mathrm{#1}}
\newcommand{\GeV}{\unit{GeV}}

\newcommand{\percc}{\unit{cm}^{-3}}
\newcommand{\Msun}{M_{\odot}}

\newcommand{\msunpyr}{\Msun\unit{yr}^{-1}}

\newcommand{\JADEStwelve}{JADES-GS-z12-0~}
\newcommand{\JADESeleven}{JADES-GS-z11-0~}

\newcommand{\JADESzthirteen}{JADES-GS-z13-0~}
\newcommand{\JADESfz}{JADES-GS-z14-0~}
\newcommand{\JADESfo}{JADES-GS-z14-1~}
\firstpage{1} 
\pubvolume{1}
\issuenum{1}
\articlenumber{0}
\pubyear{2026}
\copyrightyear{2025}
\externaleditor{Firstname Lastname} % More than 1 editor, please add `` and '' before the last editor name
\datereceived{11 November 2025} 
\daterevised{9 December 2025} % Comment out if no revised date
\dateaccepted{13 December 2025} 
\datepublished{ } 
\hreflink{https://doi.org/} % If needed use \linebreak
\Title{Supermassive Dark Stars and Their Remnants as a Possible Solution to Three Recent Cosmic Dawn Puzzles}

\Author{Cosmin Ilie $^{1,}$*\orcidA{}, Jillian Paulin $^{2}$\orcidB{},  Andreea Petric $^{3,4}$ and Katherine Freese $^{5,6,7}$\orcidD{}}
\AuthorNames{Cosmin Ilie, Jillian Paulin, Andreea Petric, and Katherine Freese}

\address{%
$^{1}$ \quad  Department of Physics and Astronomy, Colgate University, 13 Oak Dr., Hamilton, NY 13346, USA\\
$^{2}$ \quad \textls[-20]{Department of Physics and Astronomy, University of Pennsylvania, 209 S 33rd St., Philadelphia, PA 19104, USA;} jpaulin@sas.upenn.edu\\
$^{3}$ \quad Space Telescope Science Institute (STScI), 3700 San Martin Dr.,
 Baltimore, MD 21218, USA; apetric@stsci.edu\\
$^{4}$ \quad \textls[-20]{William H. Miller III Department of Physics and Astronomy, Johns Hopkins University, Baltimore, MD 21218, USA}\\
$^{5}$ \quad Weinberg Institute for Theoretical Physics, Texas Center for Cosmology and Astroparticle Physics, University of Texas, 2515 Speedway, Austin, TX 78712, USA; ktfreese@utexas.edu\\
$^{6}$ \quad The Oskar Klein Centre, Department of Physics, Stockholm University, Stockholm, Sweden\\
$^{7}$ \quad Nordic Institute for Theoretical Physics (NORDITA), AlbaNova University Centre, Hannes Alfv\'{e}ns v\"{a}g 12, Stockholm, SE-106 91, Sweden}
\corres{Correspondence: cilie@colgate.edu}

\abstract{The James Webb Space Telescope (JWST) has begun to revolutionize our view of the Cosmos. The
discovery of Blue Monsters (i.e., ultra-compact yet very bright high-z galaxies) and the Little Red Dots (i.e., very
compact dustless strong Balmer break cosmic dawn sources) pose significant challenges to pre-JWST era models of the assembly of first stars and galaxies. In addition, JWST data further strengthen the problem posed
by the origin of the supermassive black holes that power the most distant quasars observed.
Stars powered by Dark Matter annihilation (i.e., Dark Stars) can form out of primordial gas clouds during the cosmic dawn era and
subsequently might grow via accretion and become supermassive. In this paper we argue that Supermassive Dark Stars (SMDSs) offer natural solutions to the three puzzles mentioned above.}


\begin{document}

%%%%%%%%%%%%%%%%%%%%%%%%%%%%%%%%%%%%%%%%%%

\section{Introduction}

 JWST uncovered a larger than expected number of high redshift Lyman break galaxies  which are very bright yet extremely compact ($r_{eff}\lesssim 400$~pc)~\citep[e.g.][and references therein]{JWSTBLueMonsters:2025A&A...694A.286F}. If interpreted as galaxies, those ``Blue Monsters''~\cite{BlueMonsters:2023MNRAS.520.2445Z}
 are too large for their epoch ($M_{\star}\gtrsim 10^9\Msun$) and their large comoving number density~\citep{BlueMonstersNumbers:2024ApJ...965...98C,BlueMonstersHighNumber:2025ApJ} is very hard to explain with standard, pre-JWST models of the formation of the first stars and assembly of the first galaxies.   While there are uncertainties, such as the estimates of how massive single stars are~\citep{Steinhardt:2023Templates, Lapi:2024tgm}, or accurately accounting for the presence of interstellar dust~\citep{Ferrara:2022}, if this tension persists as JWST collects more and more data it would be a clear indication that our current understanding of the formation and evolution of the first stars and galaxies is, at best, incomplete.  
 
 Modifications of the power spectrum in the standard cosmological $\Lambda$CDM model are unlikely to solve the paradox posed by those distant compact very bright JWST objects, as HST would already have detected any such deviations from $\Lambda$CDM~\citep{Sabti:2023}.
At face value the JWST data implies that  the star formation efficiency for many of the galaxies at $z\gtrsim 9$ exceeds $50\%$\citep{Boylan-Kolchin:2023,RedMonsters:2024}. This in stark contrast with the observed conversion rate in the ``local'' universe, where it barely approaches $10\%$. 
This discrepancy seems to imply the need for exotic high-z stellar objects that can shine as massive as a galaxy and can accrete gas very efficiently~\citep{Iocco:2024rez}. Remarkably, NIRSpec data reveals that those ``Blue Monsters'' are almost devoid of dust~\citep{JWSTBLueMonsters:2025A&A...694A.286F}.
As we will discuss in Sec.~\ref{Sec:DS}, stars powered by Dark Matter annihilations (i.e. Dark Stars~\citep{Spolyar:2008dark}) can shine as bright as a galaxy, are in a virtually dust free environment,  and can accrete the vast majority of the gas surrounding them~\citep{Freese:2010smds,Ilie:2012,Rindler-Daller:2014uja}.

The origin of the Supermassive Black Holes (SMBHs) powering the most distant observed Quasars (QSOs) remains an open question, deepened by JWST data. The most striking such example is UHZ1, the most distant quasar observed ($z\sim 10$). JWST and Chandra data shows that UHZ1 harbors an SMBH with $M_{BH}\sim 10^7\Msun$. Such enormous Black Holes (BHs) should not have existed so early if they were seeded by regular, nuclear powered stars~\citep{Bogdan:2023UHZ1}. UHZ1 is just the tip of the iceberg when it comes to the distant quasars observed, each harboring SMBHs that should not have had time to grow as massive as they are~\citep[e.g.][]{Wang:2019,Inayoshi:2020,Lupi:2021}. If one insists on a regular nuclear powered stellar seed then long period, sustained super-Eddington accretion rates are required to explain the SMBHs powering UHZ1 and the other most distant QSOs. More likely, the data implies the need for heavy Black Hole seeds, at high redshifts~\citep{Bogdan:2023UHZ1}. Direct Collapse Black Holes (DCBHs)~\citep[e.g.][]{Loeb:1994wv,Belgman:2006,Lodato:2006hw,Natarajan:2017,barrow:2018,Whalen:2020,Inayoshi:2020}  and Dark Stars~\citep[e.g.][]{Spolyar:2008dark, Freese:2008ds, Freese:2008cap,Iocco:2008cap,Freese:2010smds}~\footnote{In the literature sometimes Dark Stars are called Pop~III.1 stars~\citep[e.g.][]{
Banik:2019, TanPopIIISMDS2025,PopIII1SMBHs:2024}} naturally generate such heavy seeds.

Lastly, JWST uncovered a completely new class of objects at the cosmic dawn era, the so called little red dots (LRDs)~\citep[e.g.][and references therein]{LDRs:2024ApJ...963..129M,LRDsCensus:2024ApJ...968...38K,LRDs:2025ApJ...991...37A}. They are extremely compact, with effective radii $\lesssim 200$~pc, and thus unlikely to be galaxies, as that would require an impossibly dense packing of stars. While there is no consensus yet as to what the LRDs are, several key, puzzling facts are known: they emit weakly in UV, and not at all in X-rays; this cannot be explained by dust obscuration, in view of the lack of re-emissions at higher wavelengths, when LDRs are observed with MIRI or ALMA~\citep{LDRsNoALMA:2025ApJ...990L..61C}. Thus, it is very unlikely that LRDs are either galaxies or Quasars. The picture that seems to emerge is that many LRDs consists of a dense gas envelope that encircle an active BH; the black hole's emitted radiation provides energy that supports this gas layer, which, in turn, has sufficient density to capture and absorb high-energy ultraviolet light and X-ray radiation~\citep[e.g.][]{LRDsBHStar:2025arXiv250316596N,LRDBHStar:2025A&A...701A.168D,LRDsBHStars:2025arXiv250709085B}. In the literature, the formation of such objects (also known as quasi-stars) has been hypothesized, in the context of the Direct Collapse Black Hole scenario, as one of the stages of the formation of SMBHs seeded by the collapse of nuclear powered Supermassive Stars~\citep{Belgman:2006}. On the other hand, SMBHs seeded by the collapse of SMDSs could, in principle lead to a very similar configuration, as the macro scale astrophysics is very similar: supermassive stars that eventually run out of fuel (nuclear for DCBHs vs DM for SMDS) or attain a sufficiently large mass to become unstable under general relativistic instabilities. When either of those two conditions is met, the supermassive star collapses and forms a SMBH, still surrounded by the remnant of the outer layers of the star and the gas that had not yet been accreted into it prior to collapse. 

When combined, the three astronomical puzzles discussed above indicate that we might be on the verge of a revolution of our understanding of the cosmic dawn era. In this paper we argue that Dark Stars (DSs) are a natural solution to {\it all three} of those tensions between data and theoretical expectations. Specifically, in Sec.~\ref{Sec:DS}, we briefly review theoretical aspects of Dark Stars, and their observational status in view of the candidates found by \cite{Ilie:2023JADES} and \cite{ilie2025spectroscopicsupermassivedarkstar}. In Sec.~\ref{sec:bluemonsters} we discuss how the SMDSs candidates already identified could explain some of the ``Blue Monsters'' observed with JWST. In Sec.~\ref{sec:SMDSToSMBHs} we demonstrate that, in addition to DCBHs, Dark Stars provide an efficient seeding mechanism for the overly massive BHs observed at high-z, and in Sec.~\ref{sec:UHZ1},  we show how Dark Star seeded BHs could have grown to power the most distant quasars observed, such as UHZ1, J0313-1806, J1342+0928, and J1007+2115. In Sec.~\ref{sec:Discussion} we conclude and further elaborate how we envision Dark Stars could be part of the mystery posed by LRDs.  In order to keep the main body of the paper as concise and accessible as possible we relegate most of the technical details to the following appendices: Appendix~\ref{sec:DM} (a brief overview of the evidence for the existence of Dark Matter), Appendix~\ref{Ap:DS} (a supplement for our Sec.~\ref{Sec:DS} on Dark Stars), Appendix~\ref{Ap:SMDSToSMBHs}, (were we expand upon the discussion of Sec.~\ref{sec:SMDSToSMBHs} regarding DCBHs and/or DSs as SMBHs seeds), and Appendix~\ref{Ap:UHz1} (where we expand upon the discussion presented in Sec.~\ref{sec:UHZ1} regarding the mechanisms via which SMDSs or their SMBH remnants can become embedded in high-z galaxies).

\section{Dark Stars (DSs): from theory to first observational hints}\label{Sec:DS}

The first stars in the universe formed more than thirteen Gyrs ago, when molecular clouds comprised of hydrogen and helium began to gravitationally collapse at the centers of DM halos~\citep[e.g.][]{Abel:2001, Barkana:2000, Bromm:2003, Yoshida:2006, OShea:2007, Yoshida:2008, Bromm:2009}. The balance between all heating and all cooling mechanisms determines the point when the runaway collapse stops, and a proto-star in hydrostatic equilibrium forms. 
 The very high DM densities, along with the poor cooling available in those pristine gas clouds, are the main reasons why DM burning stars (i.e. Dark Stars) could form~\citep{Spolyar:2008dark}. The conditions identified in \cite{Spolyar:2008dark} for the formation of DSs are expected to be met only during a relatively narrow window of redshifts, e.g. $z\in[30-10]$; as such, Dark Stars should be exceedingly rare when the Universe's age is $\gtrsim500$ Myrs, a fact confirmed by \cite{Ilie:2012}, using Hubbbke Space Telescope (HST) data. While they are composed mainly of primordial H and He (from Big Bang Nucleosynthesis), with less than $1\%$ of their mass carried by DM particles, DSs are powered by Dark Matter heating alone. 
 
 Dark Stars are expected to form at the centers of micro-DM halos, with $\sim1 $ per halo. The DM reservoir can be supplied in one of two ways. First, when the Dark Star forms, it will rely on the DM spike at the center of the micro-halo, where the progenitor DM cloud collapses. In the process, the DM densities are further enhanced, as the infall of baryons steepens the gravitational potential well, and DM orbits respond by compressing. This process can be modeled quite well via the so called Adiabatic Contraction (AC) prescription~\citep[e.g.][]{Blumenthal:1985,Freese:2008dmdens,Gnedin:2004,Gnedin:2011}.  The AC mechanism relies on gravity to pull in more DM inside the star. For DM halos that are triaxial (i.e. slightly asymmetric), the AC DM reservoir can be replenished efficiently by the large population of chaotic, centrophilic DM orbits present in such halos~\citep[e.g.][]{Valluri:2010Triaxial}.  Dark Stars powered by this extended Adiabatically Contracted (AC) DM reservoir can grow (via accretion) as massive as a million suns, and their temperatures stay below $\sim 20,000$~K~\citep{Freese:2010smds}~\footnote{If one considers 100 GeV WIMPs annihilations as their power source. For lighter/heavier mass WIMPs SMDSs would be cooler/hotter.}. Conversely, if the initial DM reservoir inside of the star is not replenished efficiently by chaotic/centrophilic orbits, then for accretion rates $\lesssim0.01\Msun/$yr the DS would have exhausted this initial AC DM reservoir by the time it reaches $\lesssim 1000\Msun$, after which it enters a contraction phase. Under the assumption of non-zero interactions between DM and protons, the contracted DS can begin to efficiently capture DM from its surroundings, which in turn can provide an additional heat source. DS powered via this DM capture mechanism can also grow to become supermassive, although they will be slightly hotter than those formed via the extended AC mechanism.
 
 The equilibrium structure of DSs can be well approximated using polytropes of variable index, ranging from $n=1.5$ (fully convective when the star is born, i.e. $M\sim 1\Msun$) to $n=3$ (radiation pressure dominated, when $M_{SMDS}\gtrsim 100\Msun$)~\citep{Freese:2008ds,Freese:2008dmdens}. Ref.~\cite{Rindler-Daller:2014uja} used the stellar evolution code \texttt{MESA} to obtain detailed models (that essentially confirmed the results using polytropes). Observable properties of dark stars can be computed from their spectra, which we modeled with \texttt{TLUSTY}~\citep{hubeny2017briefintroductoryguidetlusty} based on relevant macroscopic properties (i.e. Mass, Radius, composition of the stellar atmosphere).

While cosmological simulations have not yet been able to reproduce the Dark Star phase, they show that DM annihilations lead to a suppression of the fragmentation of the gas clouds from which the first stars form. For instance, Refs.~\cite{Ripamonti:2009xw,Smith:2012,Stacy:2014} follow the collapse of a protostellar gas cloud, while including the possible effects of $100$~GeV annihilating WIMPs. Those studies find that collapse continues up the highest, resolution limited, baryonic number densities possible in their simulations: $n_B\sim 10^{14}\percc$. However, for a $100$~GeV WIMP, a proto Dark Star, in hydrostatic and thermal equilibrium, is expected to form whenever $n_B\sim 10^{17}\percc$~\citep{Spolyar:2008dark}. Therefore, contrary to claims in the literature, those studies in fact {\it do not} show that Dark Stars cannot form. However, some of those simulations find that DM annihilations can lead to the suppression of fragmentation of the gas cloud~\citep[e.g.][]{Smith:2012,Stacy:2014}, thus removing one possible obstacle to the formation of DSs~\citep{Gondolo:2013note}.

While originally proposed as zero metallicity stars powered by WIMPs~\citep{Spolyar:2008dark}, Dark Stars could in principle be powered by a large variety of Dark Matter particles, as long as there is a mechanism to convert them to Standard Model particles efficiently. Out of the leading alternative models, Axions are unlikely to power Dark Stars. However, as shown by \cite{Wu:2022SIDMDS}, Self Interacting Dark Matter (SIDM) can also lead to formation of Dark Stars, with very similar properties to those powered by WIMPs. Ongoing preliminary work on number changing DM models (e.g. SIMPs, Co-SIMPs) shows promising results in terms of the potential for those kind of particles to lead to the formation of Dark Stars. In what follows we will restrict our attention (for simplicity) to DSs powered by 100 GeV WIMPs.  

 Dark Stars are very puffy ($R_{DS}\sim 10$~A.U.); they are also relatively cool, with temperatures ranging between a few thousands and a few tens of thousands K, depending on the DM particle mass and the formation mechanism~\citep{Freese:2010smds}. This is the main reason DSs can continue to grow almost indefinitely, as long as there is a supply of baryons to accrete from and DM to annihilate. Radiative feedback effects that would shut off accretion, and which limit the mass of  nuclear burning stars to a few hundred times the mas of the Sun ($\Msun$), are almost irrelevant for  dark stars. As such, DSs convert gas into stellar material at almost $100\%$ efficiency. This is in stark contrast with the much lower, experimentally observed, efficiency of conversion in ``local'' galaxies~\citep[e.g.][]{Leroy:2008SFR,2018Starburst}. Recent JWST data indicates that many of the most distant ``galaxies'' ever observed must have converted gas into stars at an incredible $\gtrsim 60\%$ rate~\citep{Boylan-Kolchin:2023}.  Dark Stars, as well as the supermassive stars precursors to DCBHs are the only two known mechanisms via which gas is converted into stars so efficiently.

\subsection{Dark Star Candidates Found}

\begin{table}%[tbhp]
\centering
\caption{Summary of best fit parameters for the four SMDSs spectroscopic candidates found in \cite{ilie2025spectroscopicsupermassivedarkstar}.$^{*}$ Note that we assumed, for simplicity, that Dark Stars are powered by 100 GeV WIMPs, annihilating at the canonical $\langle \sigma v\rangle=3\times 10^{-26}cm^3/s$. In future work we plan to include the mass of the DM particle as a free parameter, which could be constrained based on observable properties of SMDSs.  
% Here we use the short name for each object, i.e. we omit the JADES designation. SMDS mass$^{*}$, estimated redshift, the density of H in the nebula surrounding the SMDSs ($n_H$), the effective size of the nebula, and the S\'{e}rsic Index are listed. As \JADESfo is consistent with a point object, it was modeled as a pure SMDS, by assuming negligible $N_H$. The best fit SMDSs SEDs are plotted against NIRSpec data in Fig.~\ref{fig:SMDSsCandidates}
\label{tab:combined_properties}}
\begin{adjustwidth}{-\extralength}{0cm}
\begin{tabular}{lcccccc}
\toprule
Object Name & $\log (M_{\mathrm{SMDS}}/M_{\odot})$ & $\log(M_\star/M_\odot)$ & Redshift ($z$) & $\log (\mathrm{n_{H}/cm^{-3}})$ & $r_e$ (pc) & S\'{e}rsic Index \\
\midrule
GS-z11 & $5.803^{+0.005}_{-0.006}$ & $8.67^{+0.08}_{-0.13}$ & $11.38^{+0.001}_{-0.001}$ & $-0.04^{+0.18}_{-0.18}$ & $194.00^{+155.20}_{-155.20}$ & $4.00^{+0.05}_{-0.05}$ \\
GS-z13 & $5.715^{+0.011}_{-0.011}$ & $7.95^{+0.29}_{-0.36}$ & $13.18^{+0.05}_{-0.05}$ & $0.23^{+0.03}_{-0.03}$ & $115.00^{+12.21}_{-12.21}$ & $4.00^{+0.03}_{-0.03}$ \\
GS-z14-0 & $6.222^{+0.003}_{-0.003}$ & $8.84^{+0.09}_{-0.10}$ & $14.44^{+0.04}_{-0.04}$ & $-0.1^{+0.03}_{-0.03}$ & $440.00^{+32.59}_{-32.59}$ & $3.99^{+0.04}_{-0.04}$ \\
GS-z14-1 & $5.752^{+0.002}_{-0.002}$ & $8.00^{+0.40}_{-0.30}$ & $13.90^{+0.0003}_{-0.0003}$ & N/A & --- & --- \\\\
\bottomrule
\end{tabular}
\end{adjustwidth}
\noindent{\footnotesize{*  For comparison, in the third column we list the estimated stellar mass ($M_\star$) if the same objects are modeled as regular galaxies, with no Dark Star component.~\citep[see][Table~1]{JWSTBLueMonsters:2025A&A...694A.286F}}}
\end{table}

\begin{figure}[H]
%\isPreprints{}{% This command is only used for ``preprints''.
\begin{adjustwidth}{-\extralength}{0cm}
\centering
\subfloat[\centering]{\includegraphics[width=7.0cm]{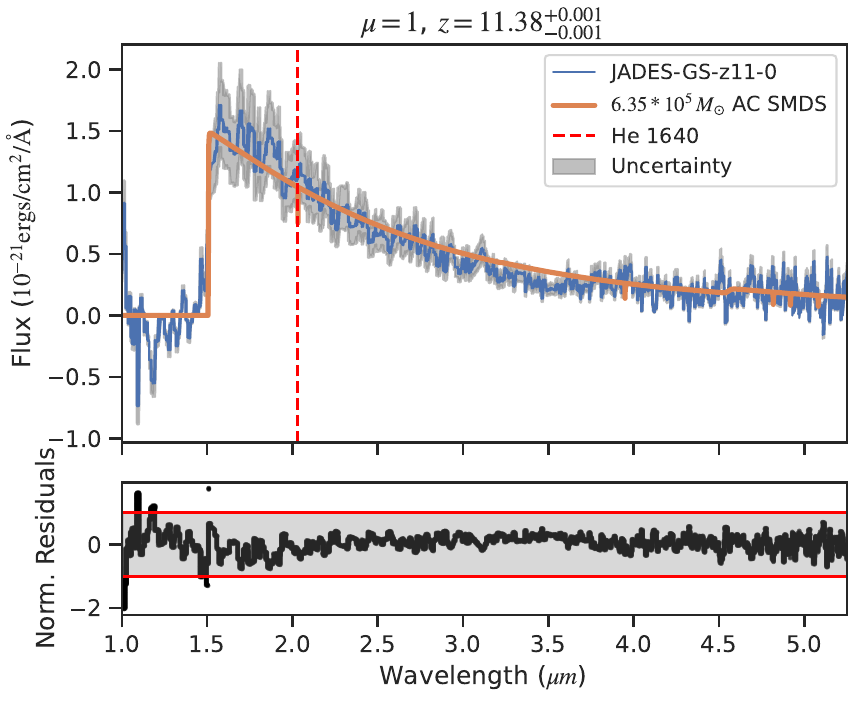}}
%\hfill
\subfloat[\centering]{\includegraphics[width=7.0cm]{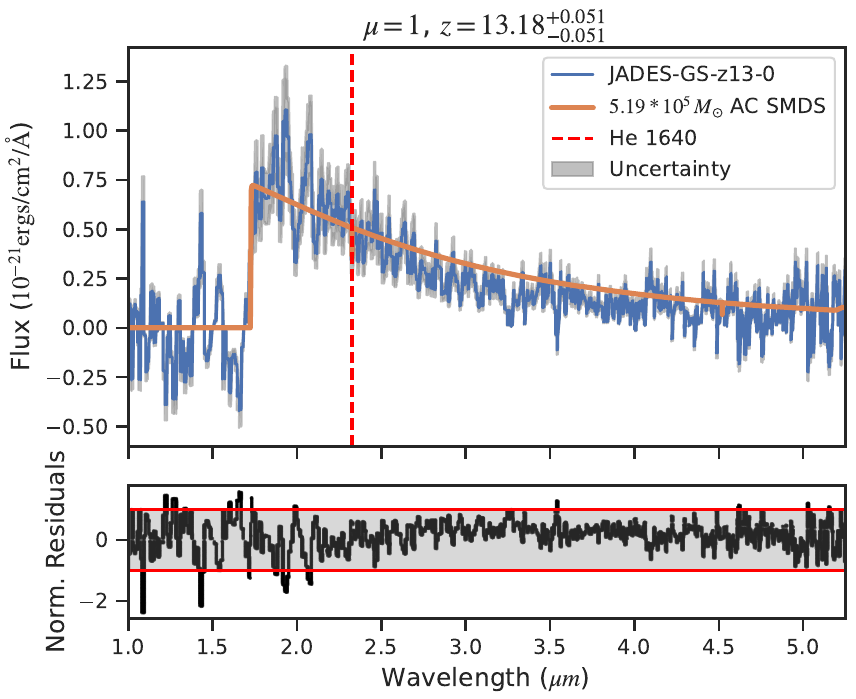}}\\
\subfloat[\centering]{\includegraphics[width=7.0cm]{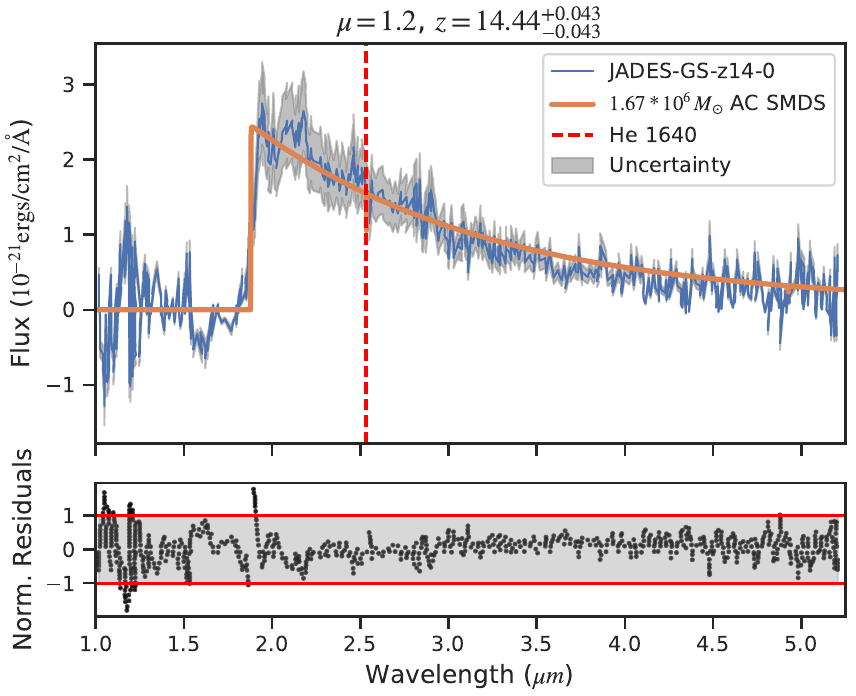}}
%\hfill
\subfloat[\centering]{\includegraphics[width=7.0cm]{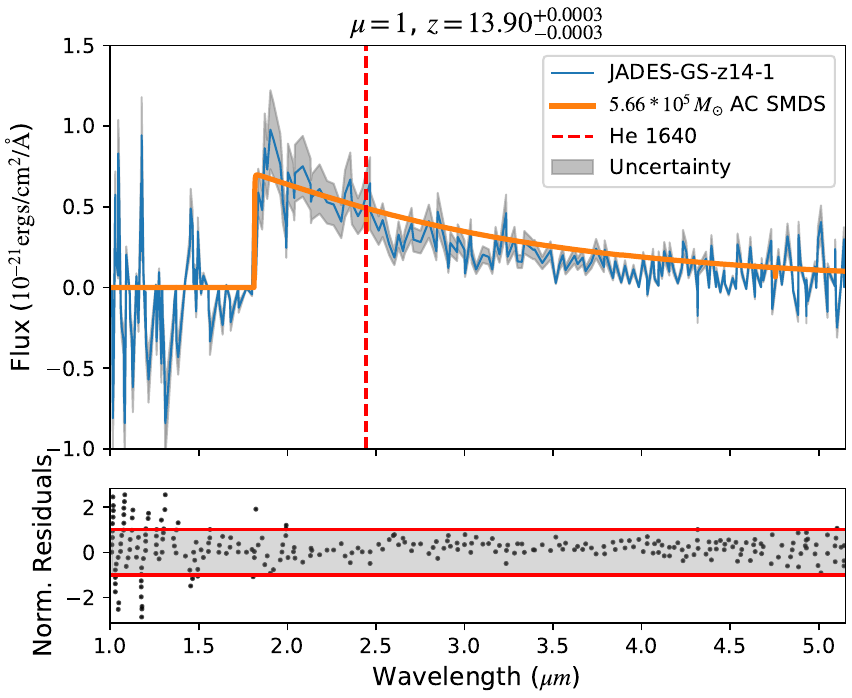}}
%\isPreprints{}{% This command is only used for ``preprints''.
\end{adjustwidth}
\caption{Spectral Emission Distribution (SED) best fits for the SMDSs candidates presented in Table~\ref{tab:combined_properties}. The data (blue line) and uncertainty band (shaded gray) plotted against our best fit Dark Star models (orange line). The red dashed line represents where He~II$\lambda1640$ absorption feature, expected only for Dark Stars, might be observed. 
%We insert a normalized residual plot ($\frac{F_{simul}-F_{\lambda}}{\sigma}$) at the bottom each plot plot which shows almost every observation falls within 1$\sigma$. 
The normalized residuals ($\frac{F_{simul}-F_{measured}} {\sigma_{measured}}$) displayed in the lower panels  of each of the SEDs show that our Supermassive Dark Star models lie consistently within 1-$\sigma$ of the NIRSpec data for each of the four objects considered. The vertical drop in the fluxes represents the Lyman break, as expected for $z\gtrsim 6$ objects due to absorption by neutral H along the line of sight~\citep{Gunn-Peterson:1965}. The vast majority of the other ``features'' in the observed spectra are actually due to noise. In the title of each plot we display the values assumed for the gravitational lensing factor ($\mu$) and the best fit values for $z_{spec}$.\label{fig:SMDSsCandidates}}
\end{figure}

\begin{figure}[!htb]
    \centering
    \includegraphics[width=0.8\linewidth]{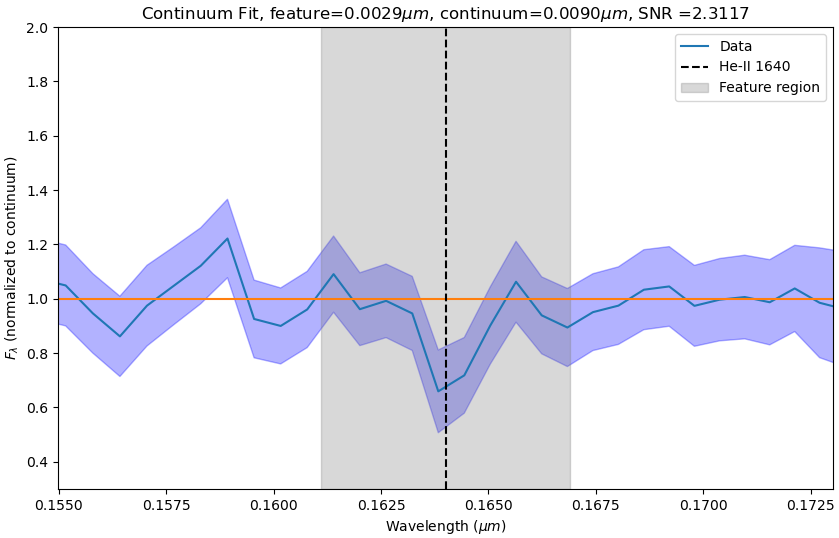}
    \caption{He~II 1640~{\AA} absorption feature identified in the spectrum of \JADESfz by \cite{ilie2025spectroscopicsupermassivedarkstar}. Here we calculate the SNR of detection  based on a polynomial fit (orange) to the observed spectrum (blue). Namely, $SNR = D_{feature}/\sigma_{cont}$, where $D_{feature}$ represents the depth of the feature below the modeled continuum, which has a scatter (noise) quantified by $\sigma$. The location of He~II line is shown in black, and the size of the feature is shaded in gray. The feature is below the continuum beyond the level of noise. Estimated $SNR\simeq2.31$.}
    \label{fig:snrcalc}
\end{figure}

JWST is sensitive enough to detect Supermassive Dark Stars, even if unlensed~\citep{Ilie:2012}, with the first ever Dark Star photometric candidates having been identified by~\citep{Ilie:2023JADES}: \JADESeleven, \JADEStwelve, and \JADESzthirteen. In a followup study, \cite{ilie2025spectroscopicsupermassivedarkstar} analyzed spectra and morphology for four of the most distant objects ever observed: \JADESeleven, \JADESzthirteen, \JADESfz, and \JADESfo and found that each of those objects are consistent with Supermassive Dark Stars powering a nebula; for a summary of values of best fit SMDSs model parameters for each of those objects see Table~\ref{tab:combined_properties}, where we use the short name for each object, i.e. we omit the JADES designation. SMDS mass ($M_{SMDS}$), estimated redshift ($z$), the density of H in the nebula surrounding the SMDSs ($n_H$), the effective size of the nebula ($r_e$), and the S\'{e}rsic Index are listed. As \JADESfo is consistent with a point object, it was modeled as a pure SMDS, by assuming negligible $n_H$. The best fit SMDSs SEDs are plotted against NIRSpec data in Fig.~\ref{fig:SMDSsCandidates}.~\footnote{For galactic fits see \cite{hainline2024searching} (Figs.~3 and 4 there) presenting fits for \JADESeleven and \JADESzthirteen) and \cite{carniani2024shining} (extended data Figs.~9 and 10) for fits of \JADESfo and \JADESfz. At the level of the continuum the available NIRSpec data cannot be used to disambiguate between a galactic and a SMDSs interpretation for those four objects.} More details regarding methodology of finding the best fit models, and explicitly showing how their morphology matches that inferred for those objects from NIRCam data can be found in \cite{ilie2025spectroscopicsupermassivedarkstar}. For \JADESfz a tentative He~II 1640~{\AA} absorption feature, a smoking gun signature for dark stars, has been identified at a $S/N\sim 2$ level by \cite{ilie2025spectroscopicsupermassivedarkstar} (see Fig.~\ref{fig:snrcalc}).

\begin{figure}[!htb]
\includegraphics[width=0.49\linewidth]{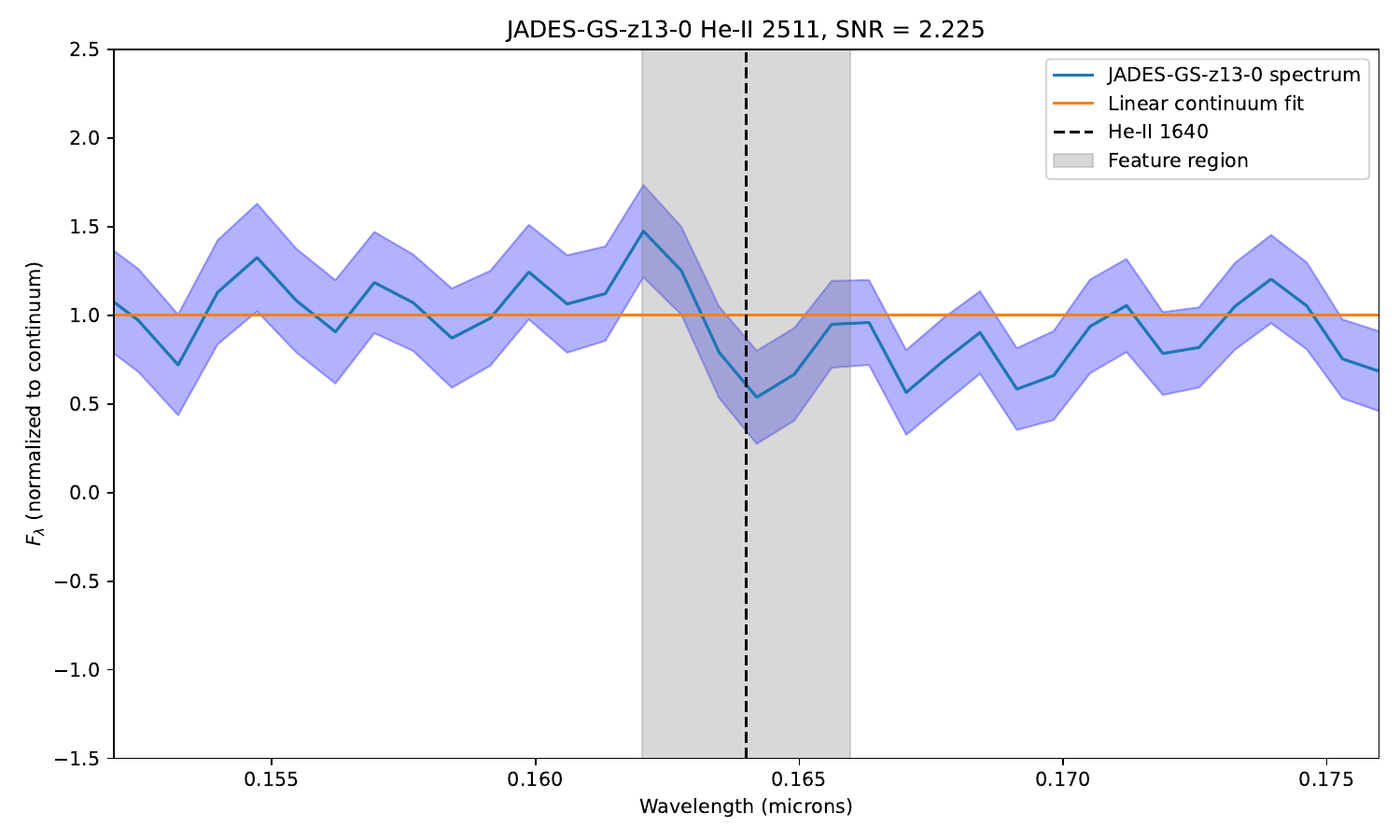}\includegraphics[width=0.49\linewidth]{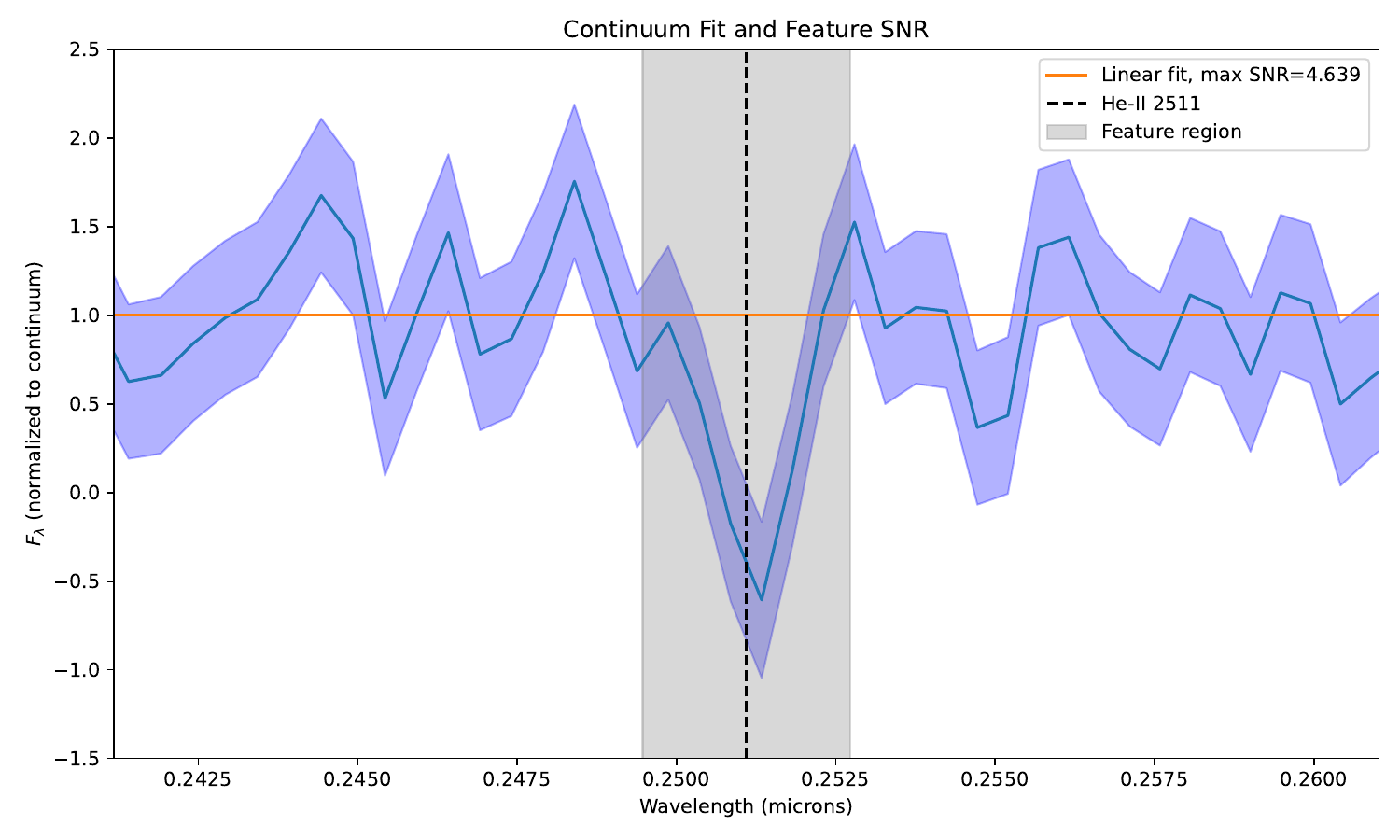}
    \caption{Absorption features identified in the spectrum of \JADESzthirteen.  Left panel: He~II 1440~{\AA} Estimated $SNR\simeq2.2$. Right panel: He~II 2511~{\AA}. Esimated $NSR\simeq 4.6$.}
    \label{fig:He2511}
\end{figure}

In Fig.~\ref{fig:He2511} we show the two He~II absoprtion features identified by us in the NIRSpec spectrum of \JADESzthirteen: a He~II~{1640\AA} ($SNR\sim 2$) absoprtion feature (left panel) and a deeper ($SNR\sim 4$) He~II 2511~{\AA} absorption feature ($SNR\simeq 4.5$).  He~II 2511~{\AA} is rarely used as a diagnostic tool in astrophysics, as one typically relies on the He~II 1640~{\AA} line instead, which is expected to be stronger, as explained below. The He~II$\lambda2511$ line is given by a transition from $n=3\to 7$ of the single electron in He~II, whereas the He~II$\lambda1640$ absorption is due to a transition from $n=2\to 3$. In local thermodynamic equilibrium, and for the effective temperatures SMDSs (i.e. $T_{eff}\sim 20000$~K) the $n=2$ state is significantly more populated than the $n=3$ state, hence the 1640 line is expected to be stronger. However, due to a larger line opacity,  the $\lambda 1640$ transition forms high in the atmosphere where wind infilling, scattering, or velocity gradients can flatten the feature and reduce its contrast against the continuum. Moreover, in the region of the photosphere where this line forms, departure from LTE is expected to significant, as collisions are unable to maintain LTE, and photo-ionization and recombination cascades can depopulate the 
$n=2$ level. In contrast, 
the $\lambda 2511$ line is much weaker in intrinsic opacity, forms deeper in denser layers closer 
to LTE, and is largely insensitive to wind infilling, velocity gradients, or scattering.

Refs.~\cite{Zhang:2022,ilie2025spectroscopicsupermassivedarkstar,NNSMDSPhot,NNSMDSSpectra} offer further details on Dark Stars and the potential to detect them (for a brief summary see Appendix~\ref{Ap:DS}; for a review see \cite{Freese:2016dark}). As JWST continues to discover more high redshift objects, some of them lensed to high magnification, it will become possible  for 
future spectroscopic studies with the low-resolution spectrograph on JWST's Mid-Infrared Instrument (MIRI)  and the Near Infrared Spectrograph (NIRSpec) to differentiate between SMDS and galaxy interpretations~\citep{Ilie:2023JADES}.
The unambiguous discovery of even a single dark star, based on smoking gun spectral signatures such as a Helium absorption features (e.g. 1640~{\AA}, \footnote{We note here that accretion onto Direct Collapse Black Holes (DCBHs) is expected to produce strong nebular {\it emission} due to Helium-II at the same $1640~\unit{\AA}$ wavelength~\citep{Johnson:2011DCBHSEDs}.} or 2511~{\AA})  would constitute (indirect) detection of Dark Matter. Additionally, it would usher a new era in astronomy, where Dark Matter parameters estimations are extracted from observable properties of Dark Stars.

\section{Supermassive Dark stars at the core of the JWST's Blue Monsters?}\label{sec:bluemonsters}

Despite their size ($\sim 10$ AU), SMDS can appear to be significantly more extended due to the effects of nebular emission~\citep{Zackrisson:2011,ilie2025spectroscopicsupermassivedarkstar}. As such they can appear as marginally resolved objects in JWST, with effective sizes of up to $\sim 500$ parsecs (see Table~\ref{tab:combined_properties}). It is therefore possible that many of the ``Blue Monster'' objects found by JWST are in fact Dark Stars masquerading as galaxies. In fact, the four objects shown to be consistent with SMDSs in \cite{ilie2025spectroscopicsupermassivedarkstar}, as discussed in Sec.~\ref{Sec:DS}, are a sub-sample of the 15 or so NIRSpec spectroscopically confirmed $z\gtrsim 10$ Blue Monsters~\citep[e.g.][see Table~1 there]{JWSTBLueMonsters:2025A&A...694A.286F}. In a future publication we plan to analyze the data for the other 11, in terms of SMDSs or SDMSs embedded in proto-galaxies. Our expectation, based on the success rate of the SMDSs analysis for the objects we focused on so far, is that most of them would be consistent with SMDSs powering a nebula  and/or SMDSs embedded in early proto-galaxies.  Below we discuss the challenges posed by the the $z\gtrsim 10$ JWST ultra-compact, yet very bright, JWST Blue Monsters to standard models of formation of the first stars and galaxies in the Universe, and then explain how each of those problems is in fact a feature, once one adopts a Supermassive Dark Star interpretation of those objects.

First, Blue Monsters are very compact, and based on their brightness, if interpreted as galaxies, they would require $\gtrsim 10^8\Msun$ stars to be packed in a few hundred parsecs (see Table.~1 in \cite{JWSTBLueMonsters:2025A&A...694A.286F}, or for the case of \JADESeleven, \JADESzthirteen, \JADESfz, and \JADESfo Table~\ref{tab:combined_properties} here). As galaxies, this would
require an extremely dense packing of stars within a few hundred parsecs. Some of those objects are not even resolved with JWST, implying sizes $\lesssim 100$~pc, further exacerbating the problem explained above. However, if interpreted as isolated Supermassive Dark Stars, potentially powering a nebula (as done in \cite{ilie2025spectroscopicsupermassivedarkstar}) compactness is a feature of the model. 

 Second, spectroscopically confirmed $z\gtrsim10$ Blue Monsters show very low dust attenuation and dust-to-stellar mass ratios ($\xi_d$) inconsistent (by at least two orders of magnitude) with models that include standard dust production and no outflows, dominated by supernova dust production over destruction. Thus, their transparency is hard to reconcile with dust physics alone without invoking additional mechanisms~\citep{JWSTBLueMonsters:2025A&A...694A.286F}. The standard solution to this problem is the Attenuation Free Model (AFM), where dust is produced in standard amounts but is pushed to kiloparsec scales by radiation-driven outflows during super-Eddington phases~\citep[e.g.][]{BlueMonsters:2023MNRAS.520.2445Z}. Conversely, \citep{JWSTBLueMonsters:2025A&A...694A.286F} tested a “dust-free” alternative by exploring whether reduced net SN dust yield and/or enhanced interstellar destruction can lower dust enough to match observations. Even allowing wide parameter variations, a no-outflow model cannot reproduce the very low dust-to-stellar mass ratios implied by JWST spectra. However, SMDSs are perfect inhibitors of dust production. As they convert large amounts of gas into a single supermassive star which eventually collapses to a SMBH, i.e. leaving no dust behind. Therefore, when compared to the standard AFM solution, a Dark Star interpretation of the Blue Monsters does not require super-Eddington phases capable of pushing dust to kilo-parsec scales. Again, what is observed and apparently a puzzle, is just a feature of Supermassive Dark Stars. 

Lastly, if interpreted as galaxies, those Blue Monsters have to have converted gas to stars at rates sometimes exceeding 50\%~\citep[e.g.][]{Boylan-Kolchin:2023}. This is in stark contrast with expectations based on numerical simulations of the formation of the first stars and galaxies, and it exceeds the conversion rate observed in the ``local'' universe by a factor of five or more. This problem can be quantified in terms of the specific star formation rate (SSFR), which, for the spectroscopically confirmed JWST galaxies at $z\gtrsim 10$ is remarkably high~\citep[e.g.][]{BlueMonsters:2023MNRAS.520.2445Z,BlueMonstersNumbers:2024ApJ...965...98C,JWSTBLueMonsters:2025A&A...694A.286F}. On the other hand, a SMDSs interpretation can alleviate this problem in two ways. First, SMDSs shine almost at the Eddington limit, i.e. attaining the lowest theoretical limit for the mass to light ratio of a a system. This is evidenced, for example in Table~\ref{tab:combined_properties}, where the stellar mass ($M_\star$) required in order to explain the four Blue Monsters discussed there as galaxies is two-three orders of magnitude higher than the mass of the corresponding SMDSs model ($M_{SMDS}$) for each object. Therefore, within a SMDSs interpretation, the inferred star formation rates would be much lower than those for obtained from the same $z\gtrsim 10$ JWST data, under a galaxy interpretation. Moreover, SMDSs are cool (i.e. $T_{eff}\lesssim 50,000~K$), and, as such, they can accrete the baryons surrounding them almost indefinitely~\citep{Freese:2010smds}. Whenever a SMDSs runs out of DM fuel, or if it grows sufficiently massive, and therefore triggers general relativistic instabilities, it will undergo a core collapse.~\footnote{In \cite{Freese2025_DarkStarInstability} we find that the collapse is triggered, depending on the mass of the WIMP powering a SMDS, somewhere in the $M_{SMDS}\sim10^5-10^6\Msun$ range.} As we shall see in the next section, this provides a natural mechanism for seeding SMBHs.

\section{Two kinds of ``heavy seeds:" Supermassive Dark Stars and DCBHs}\label{sec:SMDSToSMBHs}

The outcome of the collapse of SMDSs is simple: it leads directly to Supermassive Black Holes (SMBHs). As we will soon demonstrate, Dark Stars could  have seeded the SMBHs powering the four most distant quasars observed: UHZ1, J0313-1806, J1342+0928, and J1007+2115. As alluded to in the Introduction, those objects, and other $z\gtrsim6$ observed Quasars, point towards the existence of Oversized Black Holes (OBHs). Those are BHs that are too massive to have grown to their inferred mass if seeded by regular stars that accrete at sub-Eddington rates. 

Direct Collapse Black Holes (DCBHs)~\cite[e.g.][]{Loeb:1994wv,Belgman:2006,Lodato:2006hw,Natarajan:2017,barrow:2018,Whalen:2020,Inayoshi:2020}  are a well known mechanism of seeding OBHs. This theoretical scenario relies on nuclear burning stars entering a supergiant cool phase whenever the accretion rate exceeds $0.1~\msunpyr$~\citep[e.g.][]{Hosokawa_2012}. Thus, a very massive nuclear burning star can be born. One of the drawbacks of this mechanism is how unlikely the conditions to form a DCBH are~\citep{Natarajan:2017,DCBHsDensity}, as they require the need for a very efficient companion star forming Dark Mater halo in {\it close} proximity in order to destroy molecular hydrogen clouds inside the DCBH host halo~\citep{Belgman:2006,Kiyuna_2023}. In contrast, SMDS are able to form over a wider variety of redshifts, and a sizable fraction of DM halos would be able to seed SMDS~\citep[e.g.][]{Ilie:2012,Banik:2019}. For more details see Appendix~\ref{Ap:SMDSToSMBHs}.

Following ~\cite{Bogdan:2023UHZ1} we will focus on UHZ1, a JWST galaxy at $z\sim 10$, harboring a very bright quasar ($L_{bolo}\sim 5\times 10^{45}~\unit{erg\, s^{-1}}$). UHZ1 is the most striking example of the puzzle described above, in view of its extreme distance ($z\sim 10$) and inferred black hole mass ($M_{BH}\sim 10^7\Msun$). \cite{Bogdan:2023UHZ1,Natarajan:2023UHZ1} conclude that UHZ1 is the best evidence so far for the DCBH scenario. Here, we propose an alternative interpretation, by showing that the SMBH powering the X-ray observed spectrum of UHZ1 could be the BH remnant of a Supermassive Dark Star.

\begin{figure}[!htb]
\includegraphics[width=0.8\textwidth]{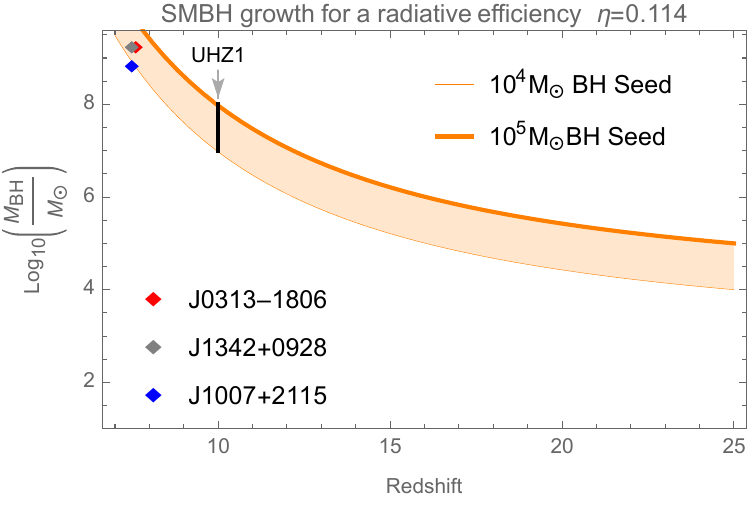}
\caption{BHs with masses between $10^4$ and $10^5 \Msun$, generated at $z\simeq 25$ and growing at the Eddington rate (tan shaded band), can explain the  mass of UHZ1 (Solid black line) and the three previously known highest redshift quasars (denoted by diamond symbols at $z\sim 7.5$). A radiative efficiency of the accretion process $\eta= 0.114$ is needed to reproduce the growth curves from Fig.~4 of \cite{Bogdan:2023UHZ1} (shaded tan band in this figure).}
\label{fig:DCBHGrowth}
\end{figure}

 Below we summarize the main evidence for the ``heavy seed'' hypothesis as an explanation for UHZ1~\citep{Bogdan:2023UHZ1,Natarajan:2023UHZ1}. In Fig.~\ref{fig:DCBHGrowth} we plot, following \cite{Bogdan:2023UHZ1}, the growth curves (tan band) for hypothetical supermassive black holes that have been seeded at $z\simeq 25$ with masses in the expected DCBH initial mass range: $10^{4}-10^{5} \Msun$. As the tan band in Fig.~\ref{fig:DCBHGrowth} passes through the inferred BH mass for UHZ1, the authors of \cite{Natarajan:2023UHZ1} claim that this is the best evidence so far for the existence of DCBHs. Note that SMDSs could collapse to BHs of the same mass range at the same redshift. Therefore, we argue that in fact UHZ1 is in fact the best evidence so far for the need of ``heavy seed'' Black Holes at high redshifts, rather than for the DCBH scenario itself. In order to generate the growth curves (i.e. the orange band in the figure) we solve:

\be\label{Eq:EddGrowth}
\dot{M}=4\times 10^{-8}\times \left(\frac{0.057}{\eta}\right) \times\frac{M}{\Msun}\times\Msun\unit{yr^{-1}}.
\ee
Eq.~\ref{Eq:EddGrowth} represents the growth rate of a quasar that shines due to accretion at the Eddington limit ($L_E=1.3\times 10^{38}(M/\Msun)~\unit{erg\,s^{-1}}$), with $\eta$ being the efficiency of the accretion process:

$$
L=\eta\dot{M}c^2
$$
Typical expected values for this parameter are $\eta\sim 0.1$, with the fiducial value of $0.057$ corresponding to accretion onto non-rotating BHs~\citep{Bardeen:1970BHs}, while for maximally rotating BHs $\eta=0.43$~\citep{Bardeen:1970BHs}. Note that if we were to use $\eta=0.057$ in Fig.~\ref{fig:DCBHGrowth}, the curves would have overshot the mass of UHZ1 and the $z\simeq 7.5$ quasars by orders of magnitude. We find that for $\eta= 0.114$ we are able to reproduce the growth rates of Fig.~4 from \cite{Bogdan:2023UHZ1}. Note that even a small change in $\eta$ would imply a significant change in the estimated BH mass at $z\sim 10$ or below. Therefore, quite a large degree of fine-tuning is necessary to explain UHZ1 and the other three $z\simeq 7.5$ quasars~\footnote{The mass estimates of J0313-1806~\citep{J0313}, J1342+0928~\citep{J1342+0928}, and J1007+2115~\citep{Yang2020} are much more accurate than that of UHZ1~\citep{Bogdan:2023UHZ1}, for which no BLR emission is detected.} in terms of DCBHs formed at $z\simeq 25$. However, the degeneracy between $\eta$ and the DCBH formation redshift alleviates this problem.

As discussed in Sec.~\ref{Sec:DS}, Dark Stars are expected to form at the centers of DM minihalos at redshifts $z_{form}\in[30,10]$. They are extremely puffy, with radii of the order of a few A.U. Their initial masses are  $\sim 1\Msun$ and they grow via accretion for as long as they are powered by DM annihilation and for as long as there is material to accrete~\citep{Freese:2010smds}. The accretion rate can be modeled with (constant) values typically ranging between $10^{-3}$ and $10^{-1} \msunpyr$. In previous studies of dark stars~\citep{Spolyar:2009,Freese:2010smds} the physically motivated variable accretion rates of~\cite{TanMcKee:2004,mckee2008formation} were used as well, confirming results based on constant accretion rates, as long as the Dark Star remains cooler than $T_{eff}\sim 50,000$~K, which is the case for all the SMDSs models studied so far in the literature. It is remarkable that the evolution along the HR diagram, and therefore the observable properties of Dark Stars powered by Adiabatically Contracted DM are insensitive to the accretion rate. Note that even for the highest constant accretion rate considered here ($10^{-1}\msunpyr$), the luminosity due to accretion onto a DS remains sub-Eddington, due primarily to the large radii of those objects.

  \begin{figure}[!htb]
\includegraphics[width=0.8\textwidth]{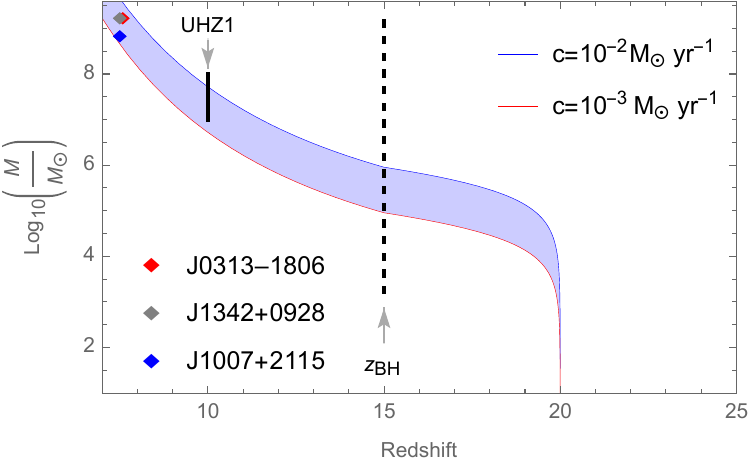}
\caption{SMBHs of: UHZ1, J0313–1806, J1342+0928, and J1007+2115 seeded by Dark Stars. The DSs is forming at $z_{form}$  and grows via accretion at a constant rate until it collapses to a BH at a redshift labeled by $z_{BH}$. The Dark Star phase is depicted by the shaded blue region to the right of the vertical dashed line at $z=z_{BH}$. The emerging BH is assumed to grow at the Eddington accretion limit (blue-shaded region to the left of $z=z_{BH}$). To be concrete we chose $z_{form}=20$, $z_{BH}=15$, and the DS mass accretion rate ranging in the conservative range $c\in[10^{-3},10^{-2}]\msunpyr$.}
\label{fig:SMDSSMBH}
\end{figure}

\begin{figure}[!htb]
\includegraphics[width=0.48\textwidth]{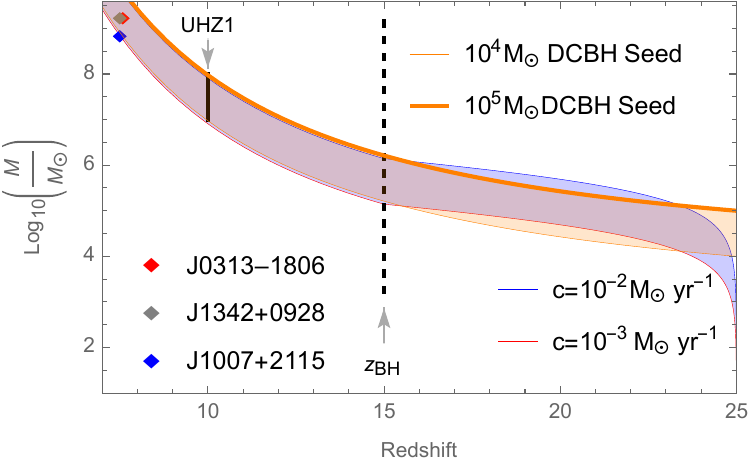}
\includegraphics[width=0.48\textwidth]{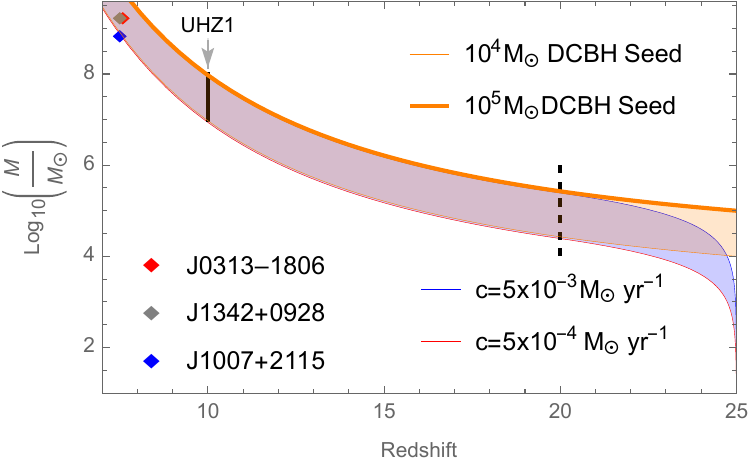}
\caption{Degeneracy between the Dark Stars (blue band) and the DCBH (tan band) solutions to the high redshift SMBHs puzzle (such as UHZ1). For both the left and right panel we chose the formation redshift for the DS to be $z_{form}=25$, whereas the redshift at which the DS collapses to a BH ($z_{BH}$; denoted by dashed line) is different: 15 (left panel) vs. 20 (right panel).  The growth of the DCBH Pop~III stellar seed is not shown in this figure.}
\label{fig:DCBHSMDSDeg}
\end{figure}

Fig.~\ref{fig:SMDSSMBH} demonstrates that Dark Stars can provide a solution to the mystery presented by the four most distant observed quasars. A DS is born at a redshift $z_{form}$, and it subsequently grows at a constant accretion rate $c$ until it collapses to a SMBH, at a redshift $z_{BH}$. The collapse can be triggered in one of three ways: i. the DS burns its DM fuel ii. the DS is dislodged from the high DM density location via mergers, or iii. The SMDSs attains sufficiently high mass ($\sim 10^5-10^6\Msun$) and it suffers a collapse due to general relativistic instabilities. The SMBH seeded via the collapse of a DS continues to grow via accretion, possibly attaining masses in excess of $10^9\Msun$ by $z\simeq 7.5$. For concreteness we chose the following values for our three parameters $z_{form}=20$, $z_{BH}=15$, and the accretion rate in the conservative range $c\in[10^{-3},10^{-2}]\msunpyr$. We find that the SMBHs seeded by this mechanism and growing at the Eddington rate~\footnote{To be consistent with the DCBH scenario discussed in Fig.~\ref{fig:DCBHGrowth} we chose the same value for the radiative efficiency: $\eta=0.114.$} can explain the large inferred mass for UHZ1 and the three other most distant quasars observed: J0313–1806, J1342+0928,  J1007+2115. Note that in addition to our choice for the values for $z_{form}$, $z_{BH}$ and $c$ made in Fig.~\ref{fig:SMDSSMBH} there is a wide range of values for those parameters that leads to the same predicted BH mass range at $z\lesssim 10$ (e.g. see Fig.~\ref{fig:DCBHSMDSDeg}). 

Fig.~\ref{fig:DCBHSMDSDeg} illustrates the degeneracy between the Dark Star collapse and the DCBH ``heavy seed'' scenarios. Namely, there is a  wide range of parameters for which, by $z\lesssim 10$, the predicted mass of the SMDSs will be identical for both. For example, if one chose the same formation redshift: $z_{form}=25$, for both the DS and the DCBH, there are numerous ways to adjust the redshift at which the DS collapses ($z_{BH}$) and the rate at which it accretes ($c$) such that for  $z<z_{BH}$ both solutions are degenerate. For instance, for $z_{BH}=15$ we find that whenever $c\in[10^{-3},10^{-2}]\msunpyr$ the growth curves align perfectly after the collapse of the DS to a BH (see left panel of Fig.~\ref{fig:DCBHSMDSDeg}). Similarly, for $z_{BH}=20$, one needs only half the previous accretion rate: $c\in[5\times10^{-4},5\times10^{-3}]$. This degeneracy is further increased if we consider other possible formation redshifts for the DS (see for example Fig.~\ref{fig:SMDSSMBH}).

Therefore Dark Stars, in addition to DCBHs, are a plausible solution to the mystery of the origin of the SMBHs powering the most distant observed quasars. In what follows we discuss in more detail UHZ1, for which the JWST IR data hints at a large stellar population, with a mass approximately equal to that of the SMBH.

\section{Dark Star interpretation of UHZ1 data or how SMDSs and/or their remnants can become embedded within high-z galaxies}\label{sec:UHZ1}

JWST IR data suggests that UHZ1 harbors a significant stellar population, with mass $\sim 10^7\Msun$. This is actually a feature of the DCBH interpretation for UHZ1~\citep{Bogdan:2023UHZ1,Natarajan:2023UHZ1}, since a large Pop~III stellar population is needed as a catalyst for the formation of the DCBH in a nearby satellite halo (for more details see Appendix~\ref{Ap:SMDSToSMBHs}). In turn, there are a variety of ways to explain the UHZ1 data using Dark Stars, the simplest of which is a SMBH SMDS system. However, in view of the extended morphology of the UHZ1 system this simple interpretation is unlikely. The second possible explanation of the data is in fact very similar to the one proposed by Refs.~\cite{Bogdan:2023UHZ1,Natarajan:2009}. Namely, the system is a galaxy that harbors a SMBH of $M_{SMBH}\sim 10^7\Msun$. In our scenario the seed of the Supermassive Black Hole is a Dark Star rather than a DCBH. 

\subsection{Mergers}\label{ssec:Mergers}

UHZ1's morphology is ``evocative of late stage mergers at low redshifts''~\citep{Bogdan:2023UHZ1}. As such, the merger of two DM halos is the most plausible explanation of UHZ1. We discuss this scenario below, in the context of a Dark Star (DS) seed for the SMBH powering the X-ray spectra of UHZ1. A DS seed forms in a DM halo at $z_{form}\in [15-25]$ (see Sec.~\ref{sec:SMDSToSMBHs}). Subsequently, the halo hosting the (by now) SMDS begins to merge with another halo (either dark or hosting nuclear fusion-powered stars). The SMDS, due to its large mass, may remain at the center of the new larger object from the merger and continue to grow. However, it is also possible that, as a result of the merger, sometimes the SMDS gets slightly dislodged from the high DM density at the center of the DM halos, and as such experience a sudden and destabilizing decrease in the DM heating.~\footnote{For details see Appendix~\ref{Ap:UHz1}.}

\begin{figure}[!htb]
\includegraphics[width=0.8\textwidth]{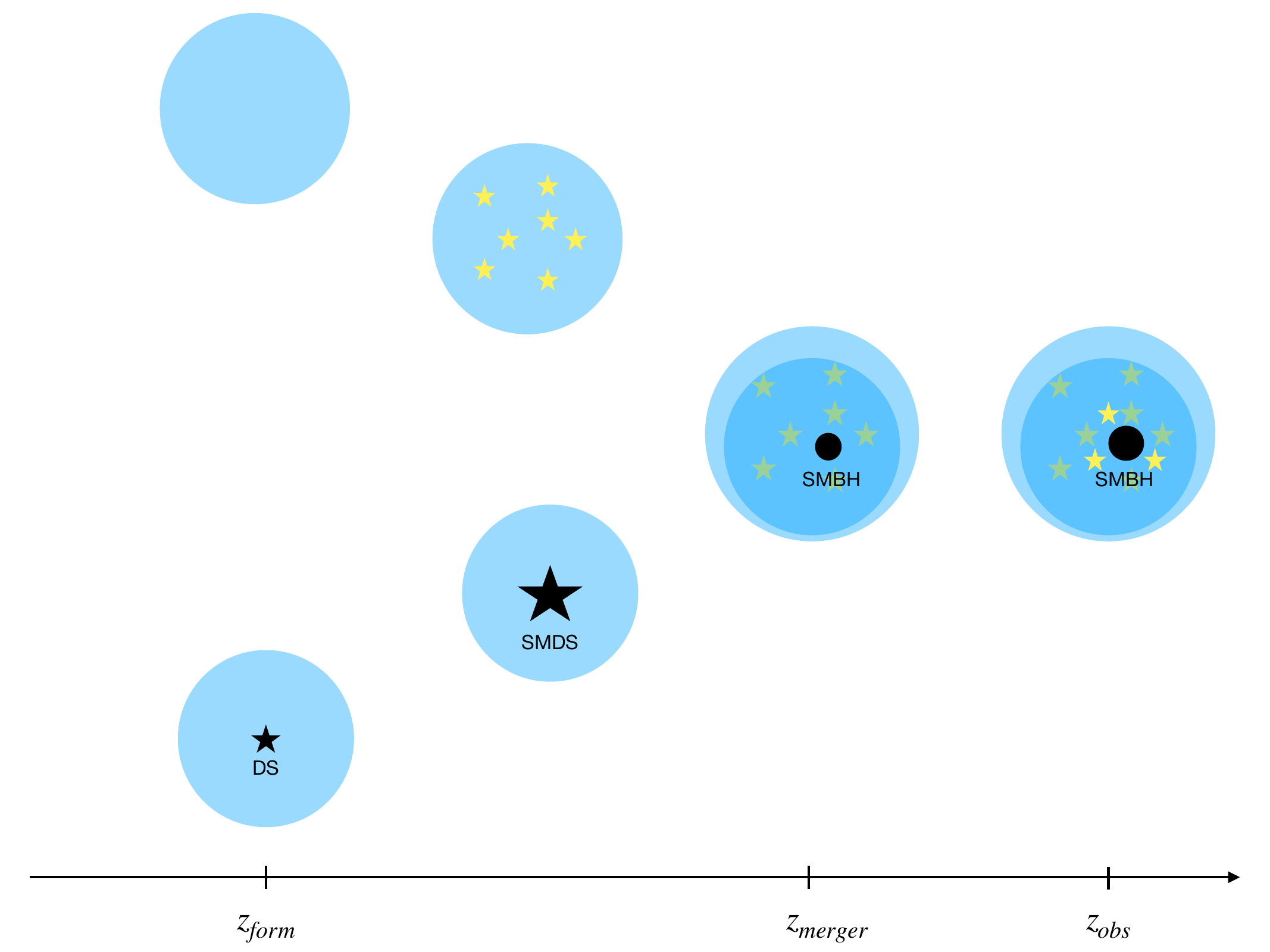}
\caption{Schematic representation of our interpretation of UHZ1 data in terms of the mergers of two DM halos. One of them hosts a Dark Star (formed at $z_{form}$) which can grow to become a SMDS.  At a redshift $z_{merger}$ this halo merges with a second DM halo. In the figure this second halo hosts regular nuclear burning stars (represented by yellow stars); or this second halo could still be dark (and a starburst happens post-merger). As soon as it runs out of DM fuel or it becomes sufficiently massive to trigger GR instabiliries, the SMDS collapses to a SMBH (black dot). The collapse can also be an outcome of mergers, whenever the SMDS would be dislodged from high DM densities at the very center of the host halo. The baryonic reservoir in its vicinity is replenished by the merger, and the SMBH can accrete efficiently until observed (at  $z_{obs}$). Moreover, ionizing radiation from the accreting SMBH triggers a starburst episode~\citep[e.g.][]{Silk:2005, Levin:2007MNRAS, carilli:2013, MF2023}(represented by bright yellow stars added for the depiction at $z_{obs}$). The growth of the DS$\to$SMDS$\to$SMBH has been modeled in Fig.~\ref{fig:DCBHSMDSDeg}, for example.}
\label{fig:Mergers}
\end{figure}

In addition to the collapse of the SMDS to a SMBH, the merger could replenish the baryonic reservoir and trigger a starburst episode in the vicinity of the SMBH~\citep[e.g.][]{Silk:2005, Levin:2007MNRAS, carilli:2013, heckman:2014, MF2023}. Thus, the SMBH can continue to grow at the Eddington rate, and be observed as the X-ray portion of the UHZ1 radiation. The stellar population embedded in those two merging halos is responsible for the majority of the IR observed data. Alternatively, UHZ1 can be explained by a halo hosting a SMBH seeded by a DS merges with another halo, leading to a starburst episode  in the newly formed system. In Fig.~\ref{fig:Mergers} we present a schematic representation of our DS model for UHz1 discussed here. 

\subsection{Same DM Halo Host}\label{ssec:OtherScenarios}

Would it be possible that all the stars within UHZ1 (including the SMDS that seeded the SMBH) have formed within the same DM halo? While unlikely (in view of the morphology of UHZ1), we briefly discuss this scenario below, as it might be relevant for other future observed high redshift quasars for which there is also a confirmed stellar counterpart. Moreover, this scenario motivates us to create, in the near future, templates for SMDSs embedded in proto-galaxies in order to fit JWST data. 

Photons in the the Lyman-Werner (LW) band ($11.2-13.6$~eV) can photo-dissociate molecular H$_2$ inside star forming H clouds, leading to a quenching of the star formation rate. \cite{Omukai:1999} shows that a Pop~III star hotter than $T_{eff}\gtrsim 54,000$~K (i.e. with mass $M\gtrsim 40\Msun$) can inhibit star formation in a region surrounding it as large as $r_{sh}\sim 1$kpc. Compared to the size of a high redshift ($z\sim 20$) DM halo, 1 kpc can be significant, or sometimes even exceed it. This would imply that star formation is almost turned off in any high redshift DM halo, as soon as the first massive Pop~III star(s) are born within it and if there are no other efficient cooling mechanisms besides H$_2$. 

Ref.~\cite{Omukai:1999} makes two key simplifying assumptions: i)  equilibrium between the dissociation and formation rates of H$_2$ within the vicinity of the star is quickly reached, and ii) the baryonic number density is constant throughout the DM halo. Accounting for the time required for the LW flux of a massive Pop~III star to photo-dissociate H$_2$ within its surroundings allows for star formation to continue efficiently in the most massive H$_2$ cooling DM halos ($M\sim 10^7\Msun$) or in the densest baryonic clumps (independent of the host halo mass)~\citep{Glover:2000}. Moreover, if assumption (ii) is replaced with a commonly used isothermal sphere profile for the gas density surrounding the star (or any other realistic profile) the estimates of the radius of influence $r_{sh}$ would be significantly reduced. 

Dark Stars are inherently cooler than Pop~III stars, so their radius of influence would be significantly smaller. A SMDS of $\sim 10^4\Msun$ emits the same LW flux as the $40\Msun$ Pop~III star considered by~\cite{Omukai:1999,Glover:2000}. Therefore it can take up to $10$ Myrs (assuming an accretion rate of $10^{-3}\msunpyr$) from the moment a DS forms until it begins to significantly suppress the star formation in its host halo. As such, it is entirely possible that Pop~III stars can form from the collapse of gas $H_2$ clumps located the same parent DM halo as a Dark Star growing towards a SMDS.  Once any of those Pop~III stars grows sufficiently massive the LW radiation will inhibit other stars from forming in their parent DM halo. Without mergers, in this scenario the total stellar mass is not expected to exceed $\sim 10^6\Msun$, in view of the fact that a higher mass would imply $M_{halo}\gtrsim 10^7\Msun$, i.e. an atomically cooled halo. We discuss this scenario next.

For DM halos with a virial temperature $T_{vir}\gtrsim 10^4$~K, atomic hydrogen cooling becomes efficient. As such, gas can cool irrespective of the flux of the LW radiation and regular star formation can continue indefinitely. Atomic Cooling Halos (ACHs) have a mass greater than $\sim 10^7\Msun$. The formation of zero metallicity clusters at the center of  ACHs has been studied in simulations, in the context of Population~III stars~\citep[e.g.][]{Kimm:2016clusers,liu:2024universal}. The growth of SMDSs at the centers of atomic cooling halos was studied using the \texttt{MESA} stellar evolution code in \cite{Rindler-Daller:2014uja}.  Since the baryonic mass within the UHZ1 system must be greater than $10^7\Msun$, in the absence of halo mergers discussed in Sec.~\ref{ssec:Mergers}, the host DM halo must have been an atomic cooling halo. Thus it is plausible (although unlikely, in view of UHZ1's distorted morphology) that all the stars within UHZ1 (including the Supermassive Dark Star we assume as the seed for the SMBH) have formed within the same DM
halo. 

 \section{Discussion and Conclusions}\label{sec:Discussion}
One of the challenges of the ``low mass seed" scenarios (in which a quasar at $z\gtrsim6$ originates from sustained Eddington-rate accretion onto an initially few stellar--mass BH seeded by a Pop~III star) is feedback. This can easily unbound the gas, thus arresting the growth of the BH. Furthermore,  growing small BHs through mergers is also challenging since the  rate is just too slow. Moreover, gravitational recoils during BH mergers can cause them to be ejected from the original halos~\citep{Whalen:2004, Haiman:2001, Haiman:2004, Whalen:2020}. Finally, recent X-ray to NIR observations of high redshift QSOs suggest that the intrinsic photon index from 0.5 to 10 keV cannot be explained by high Eddington ratios alone~\citep{farina:2022}. These lines of evidence all favor a ``heavy seed" scenario, as further demonstrated by \cite{Bogdan:2023UHZ1}. The IR luminosities, obscurations, and general observations of massive galaxies at the highest redshift suggest that the heavy seed black holes must grow without severely impacting star formation and dust production in the halos. Fueled by Dark Matter annihilations favored by the physical conditions in high redshift ($z\sim [10-25]$) Dark Matter halos, Dark Stars provide a solution to both enigmas of large galaxies and luminous QSOs observed by JWST at the highest redshifts. 

In this paper we demonstrate that even conservative accretion rates onto DSs can evolve into the SMBH powering the four most distant observed quasars: UHZ1; J0313-1806, J1342+0928, and J1007+2115 (Fig.~\ref{fig:SMDSSMBH}). We also showed that a wide range of reasonable values for (i) the formation redshifts for the Dark Star, (ii)  the accretion rate onto the DS, and (iii) the redshift when it collapses to a BH, produce the types of massive heavily obscured QSOs like UHZ1 (Figs.~\ref{fig:SMDSSMBH} and ~\ref{fig:DCBHSMDSDeg}). We further showed that subsequent mergers, similar to the late-stage galaxy mergers observed at low redshift, as suggested by the observed extended morphology of UHZ1, are compatible with a scenario in which the reservoir of baryons in the vicinity of SMDS is replenished via this merger. For UHZ1 this is the more likely SMDS interpretation, as discussed in Sec.~\ref{ssec:Mergers}. Finally, because Dark Stars are cooler than massive nuclear fusion powered stars we predict less effective feedback on the surrounding medium permitting the growth of stars in the host galaxy, even in DM halos in which atomic hydrogen cooling is not efficient. Hence our second UHZ1 DS scenario (see Sec.~\ref{ssec:OtherScenarios}): SMBH seeded by a DS embedded inside a halo hosting nuclear-powered stars.

The abundance of halos at $z\gtrsim 10$ predicted to host DCBHs is small, according to simulations~\citep[e.g.][]{DCBHsDensity}. Thus, once a larger sample of $z\gtrsim 10$ SMBHs is observed, one could place constraints on the DCBH scenario. While DSs have not yet been realized in cosmo-hydrodynamical simulations, we expect their density, at $z\gtrsim 10$ to be greater than that of DCBHs, because
\begin{enumerate}
    \item DSs can form in both ACHs and $H_2$ cooling halos~\citep{Spolyar:2008dark}.
    \item Since gas fragmentation is directly suppressed by DM annihlations~\citep{Stacy:2014}, DSs formation does not require a mechanism to destroy $H_2$.
    \item Because they are naturally puffy~\citep{Spolyar:2008dark,Spolyar:2009,Freese:2010smds}, DSs can form even when baryonic infall is low. In contrast, DCBHs can form only if the accretion rate onto the Pop~III progenitor is $\gtrsim 0.1\Msun$~\citep{Hosokawa_2012,Toyouchi_2022}.
\end{enumerate}

Gravitational waves (GWs) could be a very powerful tool to differentiate between the DCBH and the SMDS$\to$SMBH scenarios. GWs could be produced during the collapse to a BH itself (if this is asymmetric),  or during mergers of SMBHs. In both instances we expect SMBHs seeded by DSs to give stronger (and potentially detectable) signals. For instance, Ref.~\cite{Ghodla:2025fuj} shows that the GW stochastic background inferred from the Pulsar Timing Array (PTA) data~\citep{NANOGrav:2023gor} could be explained by the mergers of SMDSs, whereas DCBHs mergers could only contribute sub-dominantly. 

As discussed in Sec.~\ref{sec:bluemonsters}, SMDSs can provide a solution to the enigmatic Blue Monsters uncovered by JWST. Those are extremely compact ($r_e\lesssim 400$pc) yet very bright objects at $z\gtrsim10$, which show very little dust attenuation and, if interpreted as galaxies have to have converted gas to stars at incredibly high rates. Compactness, lack of dust attenuation, and high conversion rate of gas to stellar material are, in fact, a feature of the SMDS model, thus turning an astronomical puzzle into a potential line of evidence for SMDSs. Furthermore, We argue that Dark Stars can also be part of the solution to the Little Red Dots (LDRs) puzzle~\citep[e.g.][]{LDRs:2024ApJ...963..129M, LRDsCensus:2024ApJ...968...38K,LRDs:2025ApJ...991...37A}. Those are extremely compact ($r_{eff}\lesssim 200$~pc) sources unlikely to be galaxies due to incredibly dense stellar packing and little to no obscuration due to dust~\citep{LDRsNoALMA:2025ApJ...990L..61C}. Instead many LRDs seem to point towards the existence of SMBHs surrounded by remnant stellar layers that explain the weak UV and lack of X-ray emissions~\citep[e.g.][]{LRDsBHStar:2025arXiv250316596N,LRDBHStar:2025A&A...701A.168D,LRDsBHStars:2025arXiv250709085B}. Such objects, a.k.a. Black Hole Stars or ``quasi-stars'' are hypothesized in the DCBH scenario~\citep{Belgman:2006} and in future work we plan to investigate if such quasi-stars they can also form via the collapse of SMDSs. In addition, a large sub-sample of JWST's Little Red Dots (LRDs) are thought to harbor BHs that are significantly overmassive relative to their host galaxies~\citep{LRDsOvermassiveBHs:2025ApJ...985..169D}. Each of those Over-Massive Black Hole Galaxies (OBGs) can be easily explained by Dark Stars via the one of the two mechanisms we proposed in this paper for UHZ1: mergers (Sec.~\ref{ssec:Mergers}) or same halo (Sec.~\ref{ssec:OtherScenarios}). 

Moreover, Dark Stars could affect the re-ionization history of the Universe~\citep{Scott_2011}. As shown by~\cite{Gondolo_2022}, the inclusion of a population of DSs can naturally lead values of CMB electron scattering optical depth as low as $\tau\sim 0.05$, in agreement with the Planck measurement. This value is impossible to reconcile with only a Pop~III/II stellar population responsible for reionization, unless either very low star formation efficiencies, or strong LW feedback are invoked. In contrast, in view of their small predicted abundance, Direct Collapse Black Holes are not expected to contribute significantly to re-ionization history or to its CMB markers.

In summary, Supermassive Dark Stars can offer a solution several pressing puzzles in astronomy and astrophysics, as discussed in depth in this paper. To our knowledge, there is no other mechanism that can achieve this simultaneously.

% %%%%%%%%%%%%%%%%%%%%%%%%%%%%%%%%%%%%%%%%%%
% \section{Conclusions}

% This section is not mandatory, but can be added to the manuscript if the discussion is unusually long or complex.

% %%%%%%%%%%%%%%%%%%%%%%%%%%%%%%%%%%%%%%%%%%
% \section{Patents}

% This section is not mandatory, but may be added if there are patents resulting from the work reported in this manuscript.

%%%%%%%%%%%%%%%%%%%%%%%%%%%%%%%%%%%%%%%%%%
\vspace{6pt}

\funding{ This research was funded by the Picker Interdisciplinary Science Institute (ISI) grant number 826837.}

\acknowledgments{K.F. is grateful for support from the Jeff and Gail Kodosky Endowed Chair in Physics  at the Univ. of Texas, Austin.   K.F.  acknowledges funding from the U.S. Department of Energy, Office of Science, Office of High Energy Physics program under Award Number DE-SC0022021. K.F. acknowledges support by the Vetenskapsradet (Swedish Research Council) through contract No. 638- 2013-8993 and the Oskar Klein Centre for Cosmoparticle Physics at Stockholm University. A.P. acknowledges STScI research funding D0101.90257. C.I. acknowledges funding from Colgate University via the Research Council and the Picker Interdisciplinary Science Institute and the New York Space Grant via a faculty fellowship. CI would like to thank Sohan Ghodla for useful discussions.}

\conflictsofinterest{The authors declare no conflicts of interest.} 

%%%%%%%%%%%%%%%%%%%%%%%%%%%%%%%%%%%%%%%%%%
%% Optional

%% Only for journal Encyclopedia
%\entrylink{The Link to this entry published on the encyclopedia platform.}

\abbreviations{Abbreviations}{
The following abbreviations are used in this manuscript:
\\

\noindent 
\begin{tabular}{@{}ll}
AC & Adiabatic Contraction\\
BH & Black Hole\\
CDM & Cold Dark Matter\\
DM & Dark Matter\\
DS & Dark Star\\
HST  & Hubble Space Telescope\\
JWST & James Webb Space Telescope\\
OBH  & Over-massive Black Hole\\
OBG  & Over-massive Black Hole Galaxy\\
SMDS & Supermassive Dark Star\\
SMBH & Supermassive Black Hole\\
WIMP & Weakly Interactive Massive Particle\\
\end{tabular}
}

%%%%%%%%%%%%%%%%%%%%%%%%%%%%%%%%%%%%%%%%%%
%% Optional
\appendixtitles{yes} % Leave argument "no" if all appendix headings stay EMPTY (then no dot is printed after "Appendix A"). If the appendix sections contain a heading then change the argument to "yes".
\appendixstart
\appendix
\section[\appendixname~\thesection]{Dark Matter, a brief overview}\label{sec:DM}

Since Dark Stars rely on Dark Matter particles as a power engine, or, conversely, as they can be used as DM probes, we briefly review here the main evidence for the existence of Dark Matter (DM) and strategies for its eventual discovery. The fundamental nature of DM has remained largely elusive since Fritz Zwicky coined the term Dunkle Materie (i.e. Dark Matter) in 1933~\citep{Zwicky:1933}. Its gravitational effects have since been inferred experimentally at scales ranging from galactic~\citep[e.g.][]{Rubin:1970},  galactic cluster~\citep[e.g.][]{Natarajan:2017}, and cosmological~\citep[e.g.][]{Madhavacheril:2014,Vikram:2015,Hikage:2018,Aghanim:2018}. Moreover, DM plays a significant role in the early universe, leaving its imprint in afterglow radiation from the Big Bang, the so-called Cosmic Microwave Background (CMB) radiation. A combination of all those data leads to the emergence of the  concordance cosmological model ($\Lambda$-CDM), where the vast majority of the energy budget of the Universe is in the form of two mysterious components: Dark Energy ($70\%$) and Dark Matter (25\%). The former is responsible for the current accelerated expansion of our universe~\citep{Riess:1998SN}, while the latter assembles in large scale filamental structures, thus providing a scaffolding for the ordinary matter to assemble in galaxies and clusters. The thought that all the stars in all the galaxies in our universe form less than one percent of the total energy budget of our Universe is unsettling, perhaps even scary.

It is now firmly established that the majority of the DM in the Universe comes in the form of new, yet to be discovered particles. Amongst those, Axions and Weakly Interacting Massive Particles~(WIMPs) are perhaps the two leading candidates, in view of their natural theoretical motivations. There are three different strategies with which one can approach the hunt for Dark Matter particles: direct detection experiments (XENON~\citep[e.g.][]{Aprile:2018,Aprile:2020}, LZ~\citep[e.g.][]{LZ:2022lsv,LZ:2024psa}, PICO~\citep[e.g.][]{Amole:2019fdf}, etc), particle production in accelerators~\citep[e.g.][]{boveia2018dark} and lastly, indirect detection. The later method relies on potential observable signals of DM self-annihilating, or decaying, in various astrophysical objects or environments. As of yet, none of those methods lead to a discovery signal, although there are tantalizing hints, especially from indirect detection methods. For example, both the AMS-02 anomalous anti-proton signal, and the Fermi Gamma ray excess from the Galactic Center, can both be explained by the self annihilation of $\sim 60$~GeV WIMPs~\citep{Cholis:2019}. For both of those excesses there are also non-DM astrophysical interpretations. This is possible because of the somewhat large uncertainties, at a theoretical level, in the backgrounds for both of those signals. For reviews on Dark Matter and its theoretical or observational status the reader can consult Refs.~\citep{Freese:2017dm,arbey2021dark}.

We end this section with a non exhaustive list of objects that can act as indirect DM probes: neutron stars~\citep[e.g.][]{Bramante:2017,Baryakhtar:2017,Bell:2018pkk,Bell:2020,Ilie:2020Comment,Nguyen:2022NS}, white dwarfs~\citep[e.g.][]{Dasgupta:2019juq,Horowitz:2020axx}, Jovian planets~\citep[e.g.][]{Leane:2020wob,Croon:2023DMExoForm}, present day stars capturing and burning Dark Matter~\citep[e.g.][]{Moskalenko:2007ApJ, Scott:2009DS, John:2024thz}, and, of particular interest for this work, the first generation of stars in the Universe~\citep[e.g.][]{Freese:2008cap, Iocco:2008cap, Ilie:2020PopIII, Ilie:2021mcms}.

\section[\appendixname~\thesection]{Dark Stars }\label{Ap:DS}

In this Appendix we expand upon the material presented Sec.~\ref{Sec:DS} of the main body of the paper. The effects of Dark Matter (DM) heating on the formation of the first stars in the universe were first analyzed in Ref.~\cite{Spolyar:2008dark}, who found the conditions necessary for the formation of stars powered by annihilations of Weakly Interacting Massive Particles (WIMPs).\footnote{Note that the same WIMP models that could explain the AMS-02 and Fermi anomalies could also power Dark Stars.}  Namely: i) energy for Dark Matter annihilations can be deposited efficiently in the protostellar gas clouds, and ii) Dark Matter heating overcomes all available cooling mechanisms for the collapsing baryonic cloud. For zero metallicity clouds, such as those out of which Dark Stars could form, the cooling is relatively inefficient, and is provided by $H_2$ cooling, supplemented by $H$ line and Compton cooling. The rate, per unit volume, at which energy from DM annihilations is deposited is: 

\be\label{eq:heatingRate}
Q_{DM}=f_Q \frac{\rho_{DM}^2(r)}{m_{DM}}\langle \sigma v \rangle,
\ee
where $f_Q$ is the fraction of energy deposited, $\rho_{DM}$ is the DM density profile, $m_{DM}$ represents the DM particle mass, and  $\langle \sigma v \rangle$ is the annihilation cross section. For thermal relics, such as WIMPs, the annihilation cross section also controls the observed relic abundance. For instance, a value of $\langle \sigma v\rangle\simeq 3\times 10^{-26}cm^3/s$ leads to the observed DM relic abundance of $\Omega_{DM}\simeq 0.3$. 

As protostellar molecular clouds collapse, the gravitational potential steepens, and, as such, DM orbits will respond, leading to an enhancement of the DM densities at the center of DM halos where the first stars form. The Blumenthal~\citep{Blumenthal:1985} method of Adiabatic Contraction (AC) is used in Ref~\cite{Spolyar:2008dark}, who find that the DM density, at the edge of the baryonic core, tracks the baryonic number density ($n_B$) in the following approximate manner:

\be\label{eq:DMDens}
\rho_{DM}\simeq 5~\GeV\percc \left(\frac{n_B}{\percc}\right)^{0.81},
\ee
wheres outside the core $\rho_{DM}\propto r^{-1.9}$. Those results were subsequently confirmed (to within factors of a few) by more precise calculations in which the possibility of non circular DM orbits is included, and various initial DM profiles, in addition to the standard NFW, are explored~\citep{Freese:2008dmdens}. When compared to an initial DM profile, the DM densities can be enhanced by as much as four to five orders of magnitude via the process described above. This leads to an enhancement of 8-10 orders of magnitude of the heating rate (see Eq.~\ref{eq:heatingRate}). At the same time, as the baryonic density becomes larger, an appreciable fraction of the energy released in DM annihilation products can be deposited inside the cloud~\citep{Spolyar:2008dark}.

Ref.~\cite{Spolyar:2008dark} finds that for $100$~GeV WIMPs the DM heating rate balances all cooling mechanisms whenever the baryonic number density $n_B\sim 10^{13}\percc$. For lighter/heavier WIMPs this happens at at lower/higher values of $n_B$, in view of their more/less efficient DM heating rates (see Eq.~\ref{eq:heatingRate}). Soon after DM heating overtakes cooling, the collapse will halt, and a proto Dark Star is formed. For instance, for $100$~GeV WIMPs, this happens when $n_B\sim 10^{17}\percc$, the size of the baryonic core is of the order of $\sim 17$ A.U.s and it has a mass of $\sim0.6\Msun$. At that stage the luminosity due to DM heating is $L_{DM}\sim 140 L_{\odot}$. Therefore, the standard
picture of the formation of the first stars is drastically
modified by WIMP Dark Matter annihilations which could halt the baryonic cloud collapse well before it reaches the critical temperatures to fuse Hydrogen. 

\begin{figure}[!htb]
    \centering
    \includegraphics[width=0.8\linewidth]{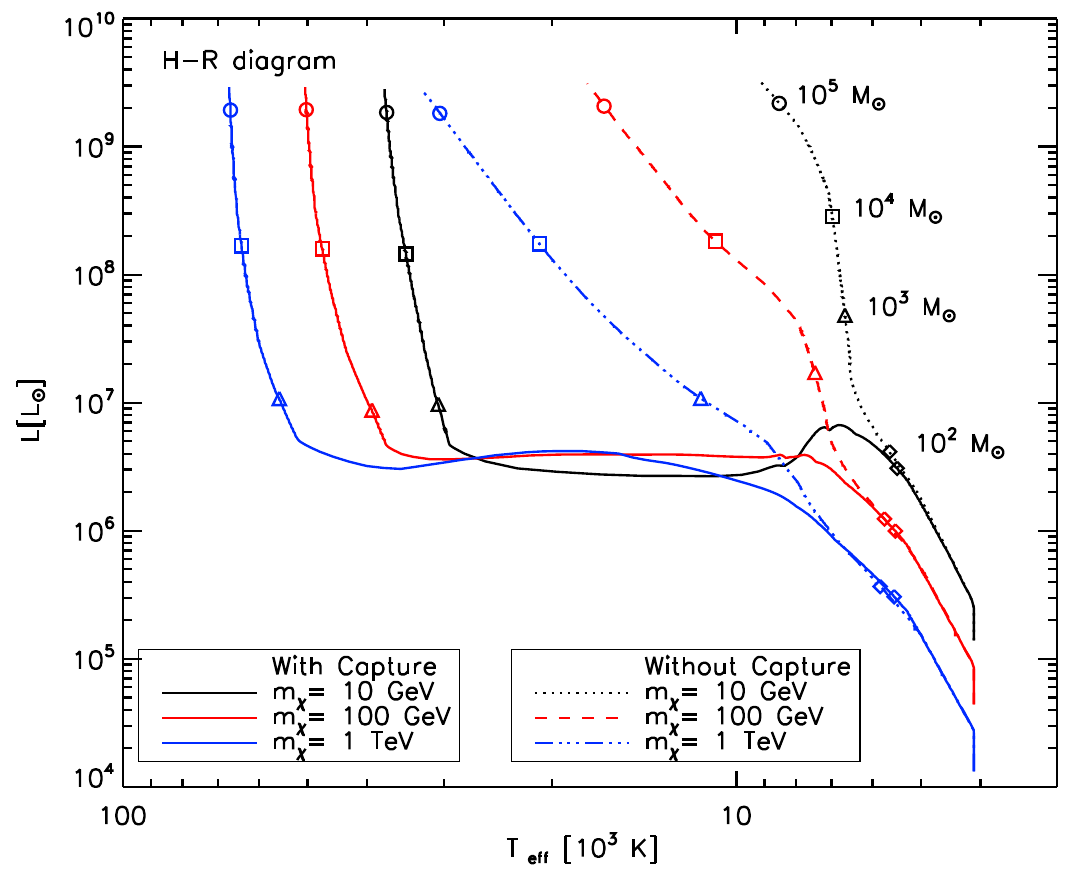}
    \caption{HR diagram for dark stars for accretion rate
  $\dot M = 10^{-3} \Msun$/yr.  and a variety of WIMP masses as
  labeled for the two cases: (i) ``without capture'' but with extended
  adiabatic contraction (dotted lines) and (ii) ``with capture''
  (solid lines).  The case with capture is for product of scattering
  cross section times ambient WIMP density $\sigma_c \bar\rho_\chi =
  10^{-39} {\rm cm}^2 \times 10^{13}$GeV/cm$^3$.  Also labeled are
  stellar masses reached by the DS on its way to becoming
  supermassive. The final DS mass was taken to be $1.5\times 10^5
  \Msun$, but SDMSs could grow in principle past this mass. (Figure adapted from \citep{Freese:2010smds}.)}
    \label{fig:HR}
\end{figure}

In Fig.~\ref{fig:HR} we reproduce the HR diagram for Dark Stars powered by WIMPs of three different masses growing at an accretion rate of $10^{-3}\msunpyr$ from \cite{Freese:2010smds}, who model using polytropes of variable index the evolution of Dark Stars from  their formation ($\sim 1\Msun$) until they become supermassive ($M\gtrsim 10^5\Msun$). Note how Dark Stars formed via the extended AC mechanism (labeled in the figure Without Capture) are cooler than those that formed via the DM capture mechanism (labeled With Capture). This is because the later underwent a Kelviln-Helmholtz contraction phase (see horizontal solid lines in the HR diagram) once they ran out of the initial AC DM reservoir. Note how lighter WIMPs are more efficient in heating DSs. This is to be expected from Eq.~\ref{eq:heatingRate}, and can be explicitly seen in the HR diagram, as at the same stellar mass, a DS powered by a lighter WIMP is puffier.   

WIMPs are not the only DM particle models that can power Dark Stars. For instance, Ref.~\cite{Wu:2022SIDMDS} identified a specific model of Self Interacting Dark Matter (SIDM) which could  satisfy
bounds from CMB, direct and indirect detection,
and, at the same time, deposit energy from annihilations into protostellar zero metallicity gas clouds efficiently. Ongoing preliminary work on light Dark Matter models that anihilate via number changin processes (e.g. SIMPs and Co-SIMPs) indicates that DSs could be powered those kind of DM particles as well. Therefore, DS are not restricted to the WIMP landscape, and we plan to explore the full DM parameter space that allows for DS to form.  

\section[\appendixname~\thesection]{Dark Stars and DCBHs: Oversized BH Seeds}\label{Ap:SMDSToSMBHs}

Here we expand upon the discussion of Sec.~\ref{sec:SMDSToSMBHs} of the main body of the paper regarding Dark Stars and Direct Collapse Black Holes as ``heavy Black Hole seeds.'' We start with abrief review of the DCBH scenario, followed by a discussion contrasting it against the Dark Stars alternative. We end with a discussion of the possible differences in the accretion rates onto Dark Stars vs those onto their BH remnants. 

 Since its initial theoretical proposal by~\cite{Belgman:2006}, the DCBH scenario has undergone several refinements and, in view of recent data has sparked significant interest. In a nutshell, cool, rapidly accreting, supermassive Population~III stars can seed Direct Collapse Black Holes. Finding environments where such rapid accretion rates are possible is challenging, but not impossible, according to simulations. For instance, in atomic cooling halos, baryonic infall rates range in the relatively wide $10^{-2}-10^{1}\Msun/yr$ range~\citep{Latif:2013pyq,Choon:2018,Patrick_2023,Toyouchi_2022}, although typically they  stay less than $1\msunpyr$. For accretion rates exceeding $\sim 0.1\msunpyr$, Pop~III stars enter a supergiant phase, leading to negligible production of ionizing radiation~\citep[e.g.][]{Hosokawa_2012}. Thus accretion can proceed without much feedback. Once they grow supermassive, Pop~III stars can even accrete at rates exceeding the atomic cooling limit ($10\msunpyr$), as shown by Ref.~\cite{Haemmerle:2019}. DCBHs are believed to form inside atomically cooled DM halos (i.e. $M_{halo}\gtrsim 10^7\Msun$) and only if there is a mechanism to destroy molecular hydrogen, or suppress its formation. Lyman-Werner (LW) photons from a nearby starburst galaxy~\citep[e.g.][]{Belgman:2006} or collisional dissociation~\citep[e.g.][]{Kiyuna_2023} are two of the most common DCBH formation channels. In addition, dynamically heated atomically cooled halos could be sites for DCBHs even if $H_2$ is not fully dissociated~\citep{Wise_2019,Li__2021,Toyouchi_2022}. In view of the very special conditions required to be met, the fraction of high redshift DM halos expected to contain a DCBH is quite rare~\citep{Natarajan:2017}.

Dark Stars are not hot enough to emit strong ionizing fluxes that would completely shut off accretion.  As such, they could potentially accrete the entirety of the gas in the surrounding cloud/disk~\citep[e.g.][]{Freese:2010smds,Banik:2019}. The Jeans scale for $H_2$ cooled gas clouds is $\sim 1000\Msun$, and it would seem that this is a reasonable guess for the upper mass of Dark Stars formed in microhalos where $H_2$ cooling is dominant. However, Dark Matter heating itself leads to a suppression of the fragmentation of the gas cloud~\citep{Stacy:2014}, so, even in the absence of any mechanism that would increase the supply of gravitationally unstable gas~\footnote{Such a mechanism is required in the DCBH scenario.} Dark Stars can easily grow to masses exceeding $1000\Msun$~\citep{Freese:2010smds}, thus providing natural seeds for massive or even supermassive BHs. Additionally, the formation of SMDSs does not imply the need for any mechanism to destroy $H_2$. In fact Dark Stars can form both in atomically and $H_2$ cooled halos~\citep{Spolyar:2009}. Moreover, when contrasted against DCBHs, SMDSs are able to form over a wider variety of redshifts, and a sizable fraction of DM halos would be able to host SMDS. In contrast, only a small fraction of high-z halos would be able to host DCBHs. Lastly, unlike the supermassive Population~III stellar precursors of DCBHs, Supermassive Dark Stars do not require very high accretion rates to remain cool. Therefore, in view of all the points above, an SMDS seed is a more flexible model for the formation of the most massive high-z quasars.

 We would like to comment below on the possible differences between the accretion rates onto the SMDSs and those onto the BHs they form. For instance, in Fig.~\ref{fig:SMDSSMBH} there are two significant differences in the growth pre- and post- BH formation. First of all, pre-BH formation the mass growth is linear with time, in view of the constant accretion rates. After the BH forms, we assume it can start accreting at its Eddington limit (for reasons explained below). As a consequence, the mass growth is exponential. Moreover, as can be seen from the discontinuity (at $z_{BH}$) in the slopes of the mass growth curves in Fig.~\ref{fig:SMDSSMBH} and the left panel of Fig.~\ref{fig:DCBHSMDSDeg}, the accretion rates onto the DS ($c$) are smaller than the accretion rate onto the $BH$, when it formed, at $z_{BH}$. Pre- BH formation, the accretion rates are limited by the stellar feedback. Since a SMDS shines at the Eddington limit, this feedback is quite strong. Once the SMDS dies, and with it the stellar feedback vanishes, the newly formed BH can now accrete the baryons surrounding it more efficiently, up to its Eddington limit.  Another possible reason for this sudden increase in the accretion rates would be due to a fresh supply of baryons in the vicinity of the newly formed BH. This could be a consequence of a merger, as in the scenario we propose in Sec.~4.1 of the main body of this paper. Lastly, we point out that the success of our model does not hinge upon an increase in the accretion rates post BH formation. For example, in Fig.6 (right panel) the stellar accretion rates smoothly match the Eddington accretion at the moment the BH forms. By slightly adjusting our parameters ($z_{BH}$, $z_{form}$, $c$) we could have also envisioned a situation where the accretion rate drops (to the Eddington limit of the BH) after the collapse of the DS to a BH, while still being able to explain UHZ-1 and the other high-z quasars considered.

\section[\appendixname~\thesection]{Collapse of SMDSs via mergers of halos}\label{Ap:UHz1}

First, any individual star moving in a field of stars experiences dynamical friction~\citep{Chandrasekhar:1943DynFric}. In this collisionless medium case, the star is decelerated due to the fluctuating gravitational field in which it moves. In our merger scenario, this would be relevant if the halo with which the DS host halo merges will have formed stars before the mergers (see Fig.~\ref{fig:Mergers}). Note that the formalism developed in~\citep{Chandrasekhar:1943DynFric} also covers the case of dynamical friction when a star is moving with respect to a Dark Matter halo, as would be the case for a SMDS host halo merging with another DM halo, irrespective of its contents. Second, a massive star moving in a gaseous medium can also experience dynamical friction, as a consequence of the gravitational attraction between the massive star and its wake in the ambient medium~\citep{Ostriker:1999DynFricGas}. In our merger scenario, if the DS hosting halo merges with a cold DM halo (in which the gas clouds have not yet collapsed to the point of forming stars), this type of dynamical friction could be relevant. Lastly, the companion halo could host a very massive central object itself, such as a Black Hole that has grown via accretion (or even another SMDS). During these types of mergers, massive central objects usually orbit each other, since they lose sufficient energy to dynamical friction\footnote{In this scenario dynamical friction can be caused by either stars, gas, or colisionless Dark Matter.} and thus become gravitationally bound~\citep[e.g.][]{Chen:2022DynFrictionMergers,Tremmel:2015DynFricMergers}.  Typically they start orbiting when separated by as much as
a few tens of kpc, lose energy (i.e. orbits shrinking) via dynamical friction, and eventually merge~\citep[e.g.][]{Tremmel:2015DynFricMergers}. Thus, during those orbits each of the two massive objects would no longer be perfectly aligned with the center of its respective DM halo, or with the center of the newly formed merged DM halo. 

We estimate below the effects of moving a SMDSs about $10 A.U.$s from the its location at center of its host DM halo. This corresponds to shifting the SMDSs by a distance comparable to its radius.  DM heating is controlled by the square of the DM density (see Eq.~\ref{eq:heatingRate}). To a good approximation the DM density in the vicinity of the SMDSs scales with distance from the DM halo center as $r^{-1.9}$~\citep{Spolyar:2008dark,Freese:2008dmdens}. Therefore, the DM heating inside of the Dark Star is reduced by a factor of $(20/10)^{-3.8}\simeq 0.07$. In other words, the DM now provides less than $10\%$ of the heating needed to sustain the SMDSs. Of course, during the two halo merger, the DM densities can be enhanced as well. Assuming those two halos are similar, this enhancement would be by a factor of $\sim 2$. Thus, in this case the DM heating is reduced by a factor of $2^2\times 0.07\simeq 0.28$, which is still quite important. 

This significant reduction in the DM heating should lead to a collapse of the SMDSs to a SMBH, since such massive objects, once the fuel is reduced, collapse directly to Black Holes. As a consequence of the merger, in addition to the collapse of the SMDS to a SMBH, an episode of rapid star formation, i.e. a starburst, is triggered in the vicinity of the SMBH~\citep[e.g.][]{Silk:2005, Levin:2007MNRAS, carilli:2013, heckman:2014, MF2023}. Moreover, the reservoir of baryons in the vicinity of the SMDS is replenished now via the merger. Thus, the SMBH can continue to grow at the Eddington rate, and be observed as the X-ray portion of the UHZ1 radiation. The stellar population embedded in those two merging halos is responsible for the majority of the IR observed data. In Fig.~\ref{fig:Mergers} we present a schematic representation of the scenario discussed above.   Or, there could already be one halo in which the SMDS has already collapsed to a SMBH, and that halo then merges with another. The merger then creates a starburst in addition to the SMBH that was already there.

%%%%%%%%%%%%%%%%%%%%%%%%%%%%%%%%%%%%%%%%%%

%%%%%%%%%%%%%%%%%%%%%%%%%%%%%%%%%%%%%%%%%%
%\isPreprints{}{% This command is only used for ``preprints''.
\begin{adjustwidth}{-\extralength}{0cm}

\reftitle{References}


\newpage
\bibliography{RefsDM,AOPref}
%=====================================
% References, variant B: internal bibliography
%=====================================

% If authors have biography, please use the format below
%\section*{Short Biography of Authors}
%\bio
%{\raisebox{-0.35cm}{\includegraphics[width=3.5cm,height=5.3cm,clip,keepaspectratio]{Definitions/author1.pdf}}}
%{\textbf{Firstname Lastname} Biography of first author}
%
%\bio
%{\raisebox{-0.35cm}{\includegraphics[width=3.5cm,height=5.3cm,clip,keepaspectratio]{Definitions/author2.jpg}}}
%{\textbf{Firstname Lastname} Biography of second author}

% For the MDPI journals use author-date citation, please follow the formatting guidelines on http://www.mdpi.com/authors/references
% To cite two works by the same author: \citeauthor{ref-journal-1a} (\citeyear{ref-journal-1a}, \citeyear{ref-journal-1b}). This produces: Whittaker (1967, 1975)
% To cite two works by the same author with specific pages: \citeauthor{ref-journal-3a} (\citeyear{ref-journal-3a}, p. 328; \citeyear{ref-journal-3b}, p.475). This produces: Wong (1999, p. 328; 2000, p. 475)

%%%%%%%%%%%%%%%%%%%%%%%%%%%%%%%%%%%%%%%%%%
%% for journal Sci
%\reviewreports{\\
%Reviewer 1 comments and authors’ response\\
%Reviewer 2 comments and authors’ response\\
%Reviewer 3 comments and authors’ response
%}
%%%%%%%%%%%%%%%%%%%%%%%%%%%%%%%%%%%%%%%%%%
\PublishersNote{}
%\isPreprints{}{% This command is only used for ``preprints''.
\end{adjustwidth}
\end{document}